\documentclass[prb,twocolumn,superscriptaddress,showpacs,amsmath,amssymb]{revtex4-1}

\usepackage{graphicx}
\usepackage{latexsym}
\usepackage{amsmath}
\usepackage{amssymb}
\usepackage{amsfonts}
\usepackage{color}
\usepackage{bm}
\usepackage{verbatim}

\definecolor{MS-color}{RGB}{128,0,128}

\usepackage{framed}
\definecolor{shadecolor}{RGB}{222,222,221}
\begin{document}

\title{
Signatures of spin-triplet Cooper pairing in the density of states of spin-textured superconductor-ferromagnet bilayers}

\author{I. V. Bobkova}
\affiliation{Institute of Solid State Physics, Chernogolovka, Moscow reg., 142432 Russia}

\author{A. M. Bobkov}
\affiliation{Institute of Solid State Physics, Chernogolovka, Moscow reg., 142432 Russia}

\author{Wolfgang Belzig}
\affiliation{Fachbereich Physik, Universit{\"a}t Konstanz, D-78457 Konstanz, Germany}

\date{\today}


\begin{abstract}
The existence of spin-triplet superconductivity in non-collinear magnetic heterostructures with superconductors is by now well established. This observation lays the foundation of superconducting spintronics with the aim to create a low-power consuming devices in order to replace the conventional electronics limited by heating. From a fundamental point of view the investigation of the structure and properties of spin-triplet Cooper pairs continues. Recently, spectroscopic evidence has shown to offer more detailed insights in the structure of triplet Cooper pairs than the supercurrent. Hence, we study here the structure of spin-triplet Cooper pairs through the density of states in bilayers of a textured magnetic insulator in proximity to a superconductor. Using quasiclassical Green function methods, we study the local density of states, both spin-resolved and spin-independent. We show that the equal-spin and mixed-spin triplet Cooper pairs leads to different spectroscopic signatures, which can be further enhanced by spin-polarized spectroscopy. Our results show the huge potential spin-polarized tunneling methods offer in characterizing unconventional superconductivity in heterostructures. 
\end{abstract}

  \maketitle

\section{Introduction}

The active development of superconducting spintronics, caloritronics and spin caloritronics raised a tremendous  interest in hybrid structures of superconductors and textured ferromagnets. The existence of spin-polarized supercurrents is ubiquitous to the spin-textured superconductor/ferromagnet (SC/FM) hybrid structures resulting from long-range spin-triplet proximity \cite{Bergeret2001, Eschrig2003, Bergeret2004, Bergeret2005b, Bergeret2005,Eschrig2015} and the following experimental measurements \cite{Keizer2006,Robinson2010,Khaire2010}. In particular, it is quite appealing to employ the spin torques generated by the dissipationless spin-polarized superconducting currents \cite{Waintal2002,Shomali2011,Zhu2016}.

On the other hand, a series of interesting
phenomena have been studied in superconductor/ferromagnet structures
with spin-split density of states (DOS), such as giant
thermoelectric \cite{Machon2013,Machon2014, Ozaeta2014, Kolenda2016, Kolenda2016_1, Giazotto2014, Giazotto2015,Rezaei2018}, thermospin effects \cite{Ozaeta2014,Linder2016,Bobkova2017}, highly efficient spin and heat valves \cite{Huertas-hernando2002,Giazotto2006,Giazotto2008,Giazotto2013,Strambini2017}, cooling at the nanoscale \cite{Giazotto2006_2,Kawabata2013} and low-temperature thermometry and development of sensitive electron thermometers \cite{Giazotto2015_2}, as well as the influence of the antiferromagnetic exchange bias \cite{Kamra2018}.

Although different F/S systems have been studied for almost three decades, investigation of the influence of magnetic proximity on the DOS in S/F layered structures with textured ferromagnets is not very much explored, contrary to DOS in the usual proximity effect \cite{Belzig1996,Gueron1996}. Recent experiments investigated tunneling in the superconducting side of spin-spiral structures \cite{DiBernado2015}. 
Furthermore, the influence of the domain structure on the position-averaged superconducting DOS in S/FI bilayer was studied experimentally and theoretically \cite{Strambini2017}.
The DOS in the ferromagnetic part of a metallic S/F bilayer with an infinitely sharp domain wall has been investigated \cite{Golubov2005}.
Characteristic signatures of equal-spin triplet pairing \cite{Diesch2018} make it possible to identify the spin-carrying supercurrent. However, the influence of particular noncollinear textures such as domain walls and other inhomogeneous ferromagnets on the local DOS practically has not yet been studied, except for the study of of Andreev resonance features near the edge of S/F bilayer containing a domain wall \cite{Bobkova2019}. The purpose of the present work is to fill in this gap.

\begin{figure}[t]
 \begin{minipage}[b]{\linewidth}
   \centerline{\includegraphics[clip=true,width=2.2in]{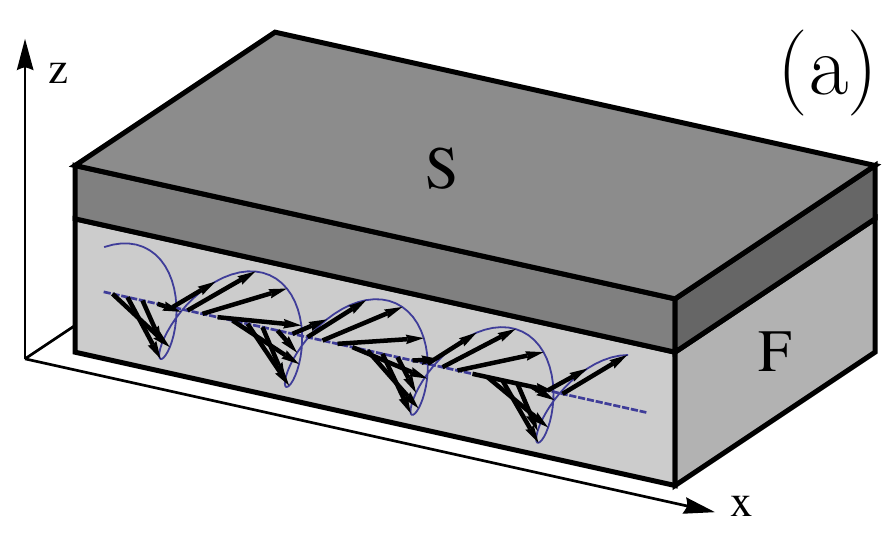}}
   \end{minipage}
   \begin{minipage}[b]{\linewidth}
   \centerline{\includegraphics[clip=true,width=2.2in]{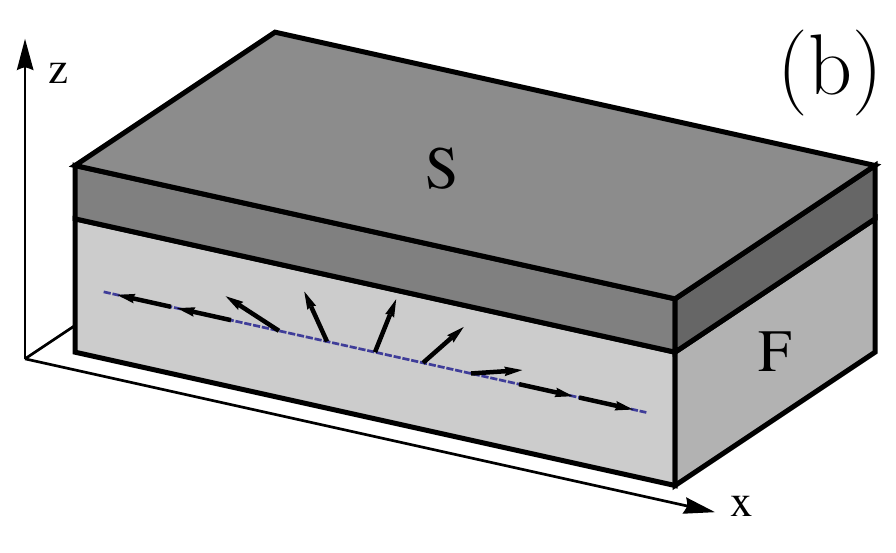}}
   \end{minipage}
   \caption{Sketch of the bilayer S/F systems considered in this work. The two examples of magnetic textures are (a) a magnetic helix and (b) a tail-to-tail domain wall.}
 \label{sketch}
 \end{figure}

In this work we investigate to two instances featuring different aspects of the spin-triplet pairs created by a magnetic textures, sketched in Fig.~\ref{sketch}. In the first part we study a magnetic helix structure in magnetic insulator in proximity to a conventional s-wave superconductor. We treat the system in the diffusive limit and show that very generically the spin-dependent density of states can be extracted from tunneling spectra. We determine the characteristic spatial and energy dependence of the spin-dependent density of states, which allows to tomographically extract the structure of the spin-triplet Cooper pairs. In the second part, we investigate a domain wall-type magnetic texture imposed in the superconducting film in the clean limit. Similar as before, the local density of states reflects that the different spin-triplet Cooper pairs have different spatial structure and one can produce in this way a one-dimensional spin-carrying band along the domain wall. Finally, we illustrate how a spin-polarized tunneling signal can be connected to the spin structure of the Cooper pairs. 

\section{DOS in diffusive spin-textured S/F bilayers}

\subsection{Model and method}

The model system that we consider is shown in Fig.~\ref{sketch}. It consists of a spin-textured ferromagnet with a spatially dependent exchange field $\bm h(\bm r)$ in contact to a spin-singlet superconductor. The ferromagnet can be a metal or an insulator. It is widely accepted in the literature that if the width of the S film $d$ is smaller than the superconducting coherence length $\xi_S$, the magnetic proximity effect, that is the influence of the adjacent ferromagnet on the S film can be described by adding the effective exchange field $\bm h_\textrm{eff}(\bm r)$ to the quasiclassical Eilenberger or Usadel equation, which we use below to treat the superconductor. This approach has been developed for metallic \cite{Bulaevskii1982,Belzig2000,Bergeret2005,Buzdin2005} as well as for insulating \cite{Tokuyasu1988,Millis1988} ferromagnets. While for the ferromagnetic insulators the magnetic proximity effect is not so simple and in general not reduced to the effective exchange only \cite{Cottet2009,Eschrig2015_2,Kamra2018}, we neglect other terms (which can be viewed as additional magnetic impurities in the superconductor) in the framework of the present study and focus on the effect of the spin texture.

In the framework of this model in the diffusive limit $l \ll \xi_S$, where $l$ is the mean free path, the S film is described by the following Usadel equation for the retarded Green's function:
\begin{eqnarray}
 -iD {\bm \nabla}\bigl( \check g {\bm \nabla} \check g \bigr) + \Bigl[ \varepsilon  \tau_z  + \bm h_\textrm{eff} \bm \sigma \tau_z   -\check \Delta, \check g \Bigr]=0,
\label{usadel_fixed}
\end{eqnarray}
where $D=v_F l/3$ is the diffusion constant with Fermi velocity $v_F$. Here and below throughout the paper the energy $\varepsilon$ contains the infinitely small imaginary part $\eta>0$, $\sigma_i$ and $\tau_i$ are Pauli matrices in spin and particle-hole spaces, respectively. $\check \Delta=\Delta \tau_+ - \Delta^* \tau_- $, where $\tau_\pm=(\tau_x\pm i\tau_y)/2$.

Further we make the spin gauge transform in Eq.~(\ref{usadel_fixed}) in order to turn to the local spin basis, where the spin quantization is aligned with the local direction of the exchange field: $\check g=U \check g_l U^\dagger$ with $U^\dagger \bm h_\textrm{eff}(\bm r)\bm \sigma U = h_\textrm{eff} \sigma_z$. In this representation the Usadel equation takes the form:
\begin{eqnarray}
 -iD \widehat {\bm \nabla}\bigl( \check g_l \widehat {\bm \nabla} \check g_l \bigr) + \Bigl[ \varepsilon  \tau_z  + h_\textrm{eff} \sigma_z \tau_z   -\check \Delta, \check g_l \Bigr]=0\,,
\label{usadel_local}
\end{eqnarray}
where the magnetic inhomogeneity enters as the spin-dependent gauge field and $\widehat {\bm \nabla} = \bm \nabla + [U^\dagger \bm \nabla U, ...]$ is the SU(2)-covariant gradient.

The quasiclassical Green's function obeys the following normalization condition $\check g^2 = \check g_l^2 = 1$. If the effective exchange field is homogeneous and the matrix Green's function $\check g$ is diagonal in spin space, it is convenient to use the so-called $\theta$-parametrization \cite{Belzig1999}, where the Green's function is parameterized by two angles $\theta_{\uparrow, \downarrow}$. This parametrization satisfies the normalization condition automatically. However, in case of textured ferromagnets the Green's function becomes non-diagonal in spin space and the parametrization via two angles fails to work. An appropriate generalization of this parametrization was proposed in Ref.~\onlinecite{Ivanov2005} and takes the form:
\begin{eqnarray}
 \check g_l = \cosh(\alpha)
 \left(
 \begin{array}{cc}
 \cosh \theta & \sinh \theta \\
 -\sinh \theta & -\cosh \theta
 \end{array}
 \right)
 + \nonumber \\
 \bm m \bm \sigma \sinh(\alpha) 
 \left(
 \begin{array}{cc}
 \sinh \theta & \cosh \theta \\
 -\cosh \theta & -\sinh \theta
 \end{array}
 \right).
\label{parametrization}
\end{eqnarray}
This parametrization satisfies the normalization condition for energy-dependent parameters $\theta(\varepsilon)$, $\alpha(\varepsilon)$ and $\bm m(\varepsilon)$, where the latter satisfies $\bm m^2=1$.

Further on, we only study 1D magnetic textures, where $\bm h_\textrm{eff}=\bm h_\textrm{eff}(x)$. Substituting Eq.~(\ref{parametrization}) into the Usadel equation Eq.~(\ref{usadel_local}), we obtain the following equations (the spatial derivatives of all the quantities are denoted by prime - $A' \equiv dA/dx$):
\begin{widetext}
\begin{eqnarray}
 iD \theta'' = 2 \cosh(\alpha) (\varepsilon  \sinh(\theta) + \Delta \cosh(\theta)) +
 2 h_\textrm{eff} m_z \sinh(\alpha) \cosh(\theta)
\label{eq:theta}
\end{eqnarray}
and
\begin{eqnarray}
 i D \Bigl( \alpha'' \bm m + \bm m' \alpha' \cosh^2 \alpha +\frac{1}{2} \bm m'' \sinh 2 \alpha -
 4 \cosh \alpha [\alpha' \cosh \alpha (\bm a \times \bm m) + \sinh \alpha (\bm a \times \bm m')] - \sinh 2 \alpha (\bm a' \times \bm m) + \nonumber \\
 2 \sinh 2 \alpha [\bm a \times (\bm a \times \bm m)]\Bigr) = 2 \sinh \alpha (\varepsilon  \cosh \theta + \Delta \sinh \theta)\bm m + 2 \cosh \alpha \sinh \theta ~ h_\textrm{eff} \bm e_z
\label{eq:m}
\end{eqnarray}
\end{widetext}
where $(\bm a)_i = -(i/2){\rm Tr}[\sigma_i U^\dagger \partial_x U]$ are the components of the gauge field $\bm a$, which can be viewed as an effective spin-orbit coupling \cite{Bergeret2013}. Eqs.~(\ref{eq:theta})-(\ref{eq:m}) together with the condition $\bm m^2=1$ are enough to find $\theta$, $\alpha$ and $\bm m$.

The Green's function in the original basis takes the form:
\begin{eqnarray}
 \check g = \cosh \alpha
 \left(
 \begin{array}{cc}
 \cosh \theta & \sinh \theta \\
 -\sinh \theta & -\cosh \theta
 \end{array}
 \right)
 + \nonumber \\
 \sinh \alpha ~ U(\bm m \bm \sigma)U^\dagger
 \left(
 \begin{array}{cc}
 \sinh \theta & \cosh \theta \\
 -\cosh \theta & -\sinh \theta
 \end{array}
 \right).
\label{g_original}
\end{eqnarray}

The DOS is calculated as
\begin{eqnarray}
N=2N_F{\rm Re}\Bigl[ \cosh \alpha \cosh \theta \Bigr]
\label{DOS}
\end{eqnarray}
and the spin-resolved DOS corresponding to the spin along the direction $\bm l$ is calculated as
\begin{eqnarray}
N_{\hat l}=N_F{\rm Re}\Bigl[ \cosh \alpha \cosh \theta + \nonumber \\
\frac{1}{2}\sinh \alpha \sinh \theta ~ \bm l_i {\rm Tr}(\sigma_i U \bm m \bm \sigma U^\dagger) \Bigr],
\label{DOS_spin_resolved}
\end{eqnarray}
where $N_F$ is the DOS per spin in the normal state of the system.

\subsection{Results: magnetic helix}

For the case of spin helix (see Fig.~\ref{sketch}(a)) the magnetization texture is described by
\begin{eqnarray}
\bm h = h (\cos \delta, \sin \delta \cos \varphi, \sin \delta \sin \varphi),
\label{h_texture}
\end{eqnarray}
where we assume that the angle $\delta$ is spatially independent and $\varphi=2\pi x/L$, where $L$ is the helix spatial period.

The spin rotation is given by
\begin{eqnarray}
\hat U = e^{-i\sigma_x \varphi/2} e^{-i\sigma_z \delta/2} e^{-i\sigma_y \pi/4}.
\label{spin_rotation}
\end{eqnarray}
In this case the gauge field is given by
\begin{eqnarray}
\bm a= \frac{\partial_x \varphi }{2 }(\sin \delta \bm e_y - \cos \delta \bm e_z)
\label{SO_helix}
\end{eqnarray}
and does not depend on coordinates. Therefore, the solution of Eqs.~(\ref{eq:theta}) and (\ref{eq:m}) and, consequently, the Green's function in the local basis also does not depend on coordinates: $\alpha'=\alpha''=0$ and $\theta' = \theta'' = 0$.

\begin{figure}[!tbh]
   \centerline{\includegraphics[clip=true,width=2.6in]{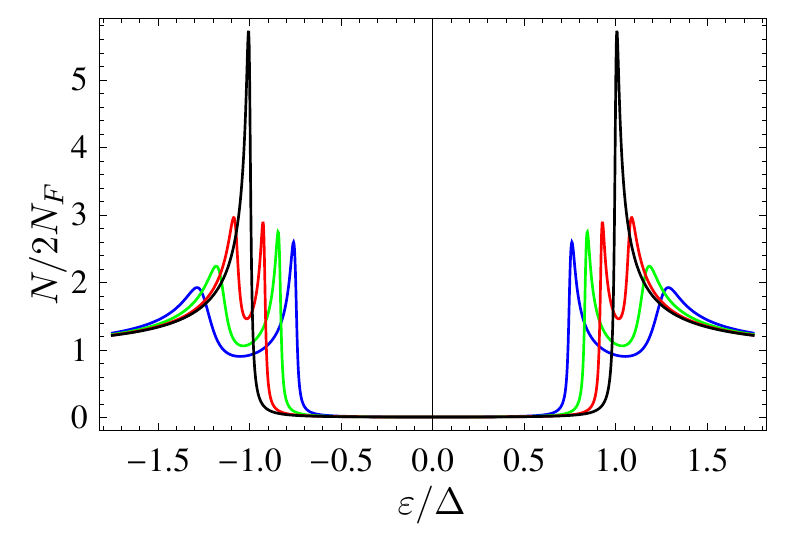}}
      \caption{DOS as a function of the quasiparticle energy for different $h_\textrm{eff}$. $\delta = 0.25 \pi$, $L/\xi_S = 5$. Different curves correspond to different $h_\textrm{eff}$: 0 (black), $0.1 \Delta $ (red), $0.2 \Delta $ (green), $0.3 \Delta $ (blue).}
 \label{DOS_h}
 \end{figure}

We solve the spatially homogeneous version of  Eqs.~(\ref{eq:theta}) and (\ref{eq:m}) numerically. For the case under consideration $\alpha(x)=const$ and $\theta(x)=const$. Therefore, in accordance with Eq.~(\ref{DOS}) the DOS is also spatially homogeneous for the case of the helix magnetic texture. Fig.~\ref{DOS_h} demonstrates the dependence of the DOS on the quasiparticle energy for different absolute values of $h_\textrm{eff}$. Here, we observe that the DOS has doubly split structure very similar to the case of spatially homogeneous exchange field. At the same time, the distance between the nearest peaks is not exactly $2h_\textrm{eff}$, as it is known for the homogeneous case. Instead, the splitting is between $h_x = h_\textrm{eff}\cos \delta$ and $h_\textrm{eff}$ and depends on the particular value of the helix spatial period, as it is discussed below. Upon increasing of $h_\textrm{eff}$ the split peaks move symmetrically around $\varepsilon = \Delta$. The peaks are smeared with respect to the homogeneous case. This smearing is due to the presence of the spin-gauge field $\bm a$, which is equivalent to the effective spin-orbit coupling. One more important feature is that for the homogeneous case the outer peaks should be higher than the inner ones. This is not always the case here: the relative height of the inner and outher peaks depend on the value of the effective spin-orbit coupling and the both cases can be realized (inner or outer peaks dominate). The fact that the inner peaks are higher due to the inhomogeneity of the ferromagnet was experimentally observed for Al/EuS hybrid structures \cite{Strambini2017} and attributed to the domain structure in EuS.

\begin{widetext}

\begin{figure}[!tbh]
 \begin{minipage}[b]{0.33\linewidth}
   \centerline{\includegraphics[clip=true,width=2.2in]{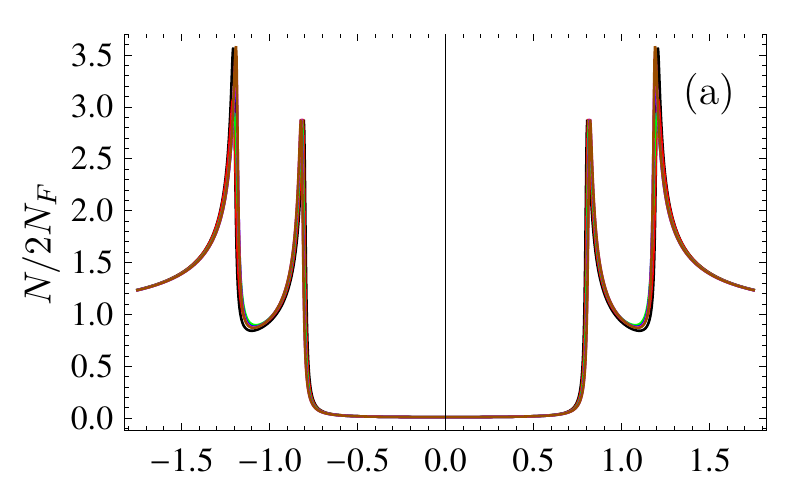}}
   \end{minipage}\hfill
   \begin{minipage}[b]{0.33\linewidth}
   \centerline{\includegraphics[clip=true,width=2.1in]{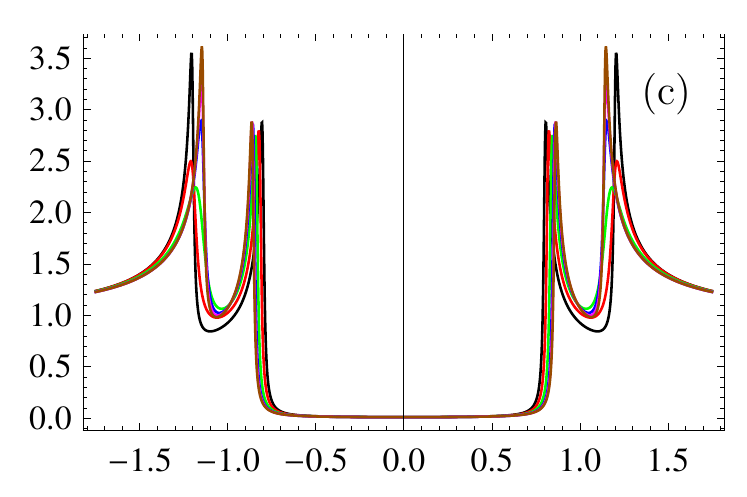}}
   \end{minipage}\hfill
   \begin{minipage}[b]{0.33\linewidth}
   \centerline{\includegraphics[clip=true,width=2.1in]{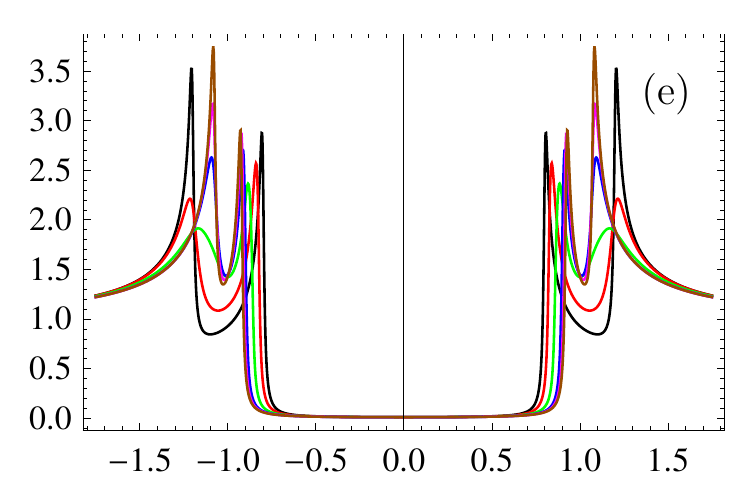}}
   \end{minipage}
   \begin{minipage}[b]{0.33\linewidth}
   \centerline{\includegraphics[clip=true,width=2.25in]{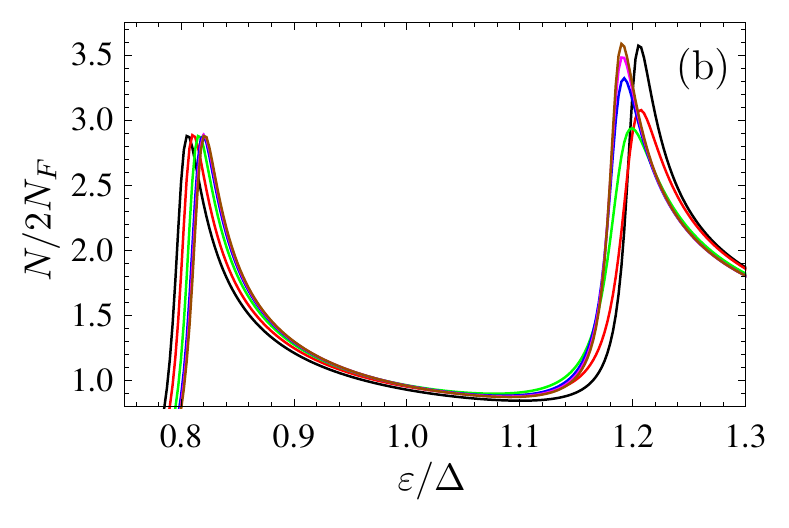}}
   \end{minipage}\hfill
   \begin{minipage}[b]{0.33\linewidth}
   \centerline{\includegraphics[clip=true,width=2.15in]{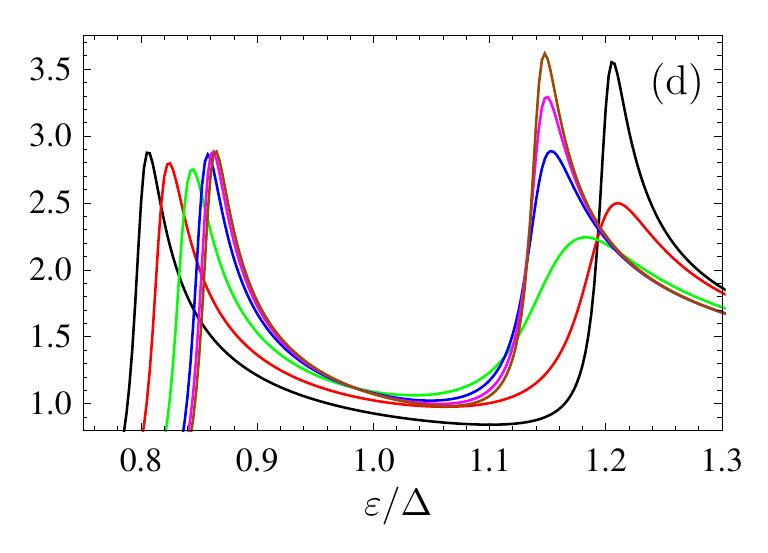}}
   \end{minipage}\hfill
   \begin{minipage}[b]{0.33\linewidth}
   \centerline{\includegraphics[clip=true,width=2.15in]{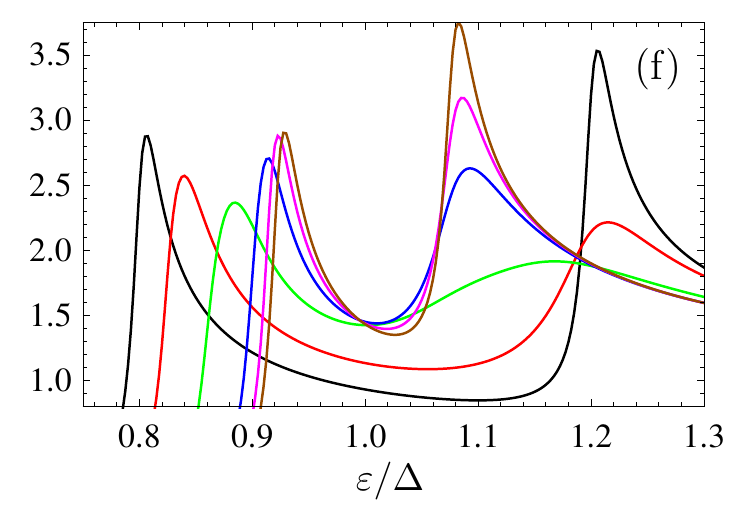}}
   \end{minipage}
   \caption{DOS as a function of quasiparticle energy for different values of the effective SO coupling, proportional to $2 \pi \xi_S/L$. Panels (a)-(b) correspond to $\delta = 0.125 \pi$; for panels (c)-(d) $\delta = 0.25 \pi$ and for panels (e)-(f) $\delta = 0.375 \pi$. The bottom row represents the energy region corresponding to the positive-energy double-split coherence peak on a large scale. $h_\textrm{eff} = 0.2 \Delta$ for all the panels. $L/\xi_S = 100$ (black), $10$ (red), $5$ (green), $2.5$ (blue), $1.5$ (pink), $0.5$ (brown).}
 \label{DOS_SO}
 \end{figure}

\end{widetext}

Now we discuss the influence of the effective spin-orbit coupling or, in other words, the value of the helix spatial period on the DOS. As it is seen from Eq.~(\ref{SO_helix}), the effective spin-orbit coupling $\bm a$ is proportional to $d\varphi/dx = 2 \pi/ L$, where $L$ is the helix period. In other words, the effective spin-orbit coupling is inversely proportional $L$. Fig.~\ref{DOS_SO} demonstrates the influence of this parameter on the DOS. The three columns of this figure differ by the value of $\delta$, viz., by the projection of the effective exchange field on the direction of the helix axis. Let us concentrate at first on the middle row of the figure. We observe that the effective SO smears the peaks only for intermediate values of $L\sim\xi_S$. It is obvious that too small effective SO ($L >> \xi_S$) should influence the peaks only slightly, but too strong SO ($L << \xi_S$) also does not cause any smearing. In this case $h_\textrm{eff}$ changes too rapidly on the length scale $\xi_S$  and the superconductor only feel the effectively homogeneous averaged exchange field $h_x$.

The effective SO not only smears the peaks, but also shifts them symmetrically with respect to $\varepsilon = \pm \Delta$. These shifts are because the averaged exchange field is reduced from $h_\textrm{eff}$ to $h_x$ upon increasing effective SO. In the limit of strong SO ($L << \xi_S$) the splitting of the DOS peaks tends to $2h_x$.

We see that the outer peaks are lower than the inner ones only for a range of intermediate values of the effective SO, where the smearing is well-pronounced, approximately corresponding to $L \sim 3-10 \xi_S$. For smaller and larger values of the effective SO the spin-texture is hardly felt by the superconductor and the DOS behaves more like in the homogeneous case.
The described above features manifest itself for the whole range of the angles $\delta$, but they are pronounced more clearly for larger values of $\delta$ because in this case the difference between $h_\textrm{eff}$ and $h_x = h_\textrm{eff}\cos \delta$ is larger. These observations are illustrated by comparing the different columns of Fig.~\ref{DOS_SO}.

As another interesting issue, we discuss now the behavior of the spin-resolved DOS. It can be calculated in the fixed spin basis according to Eq.~(\ref{DOS_spin_resolved}) and for the case of magnetic helix takes the form:
\begin{eqnarray}
N_{\pm \hat x} = \frac{N}{2} \pm {\rm Re}[\sinh \alpha \sinh \theta (m_z \cos \delta - m_y \sin \delta)],~~~~
\label{DOS_x_helix}
\end{eqnarray}
\begin{eqnarray}
N_{\pm \hat y} = \frac{N}{2} \pm g_\parallel \cos (2ax),
\label{DOS_y_helix}
\end{eqnarray}
\begin{eqnarray}
N_{\pm \hat z} = \frac{N}{2} \pm g_\parallel \sin (2ax),
\label{DOS_z_helix}
\end{eqnarray}
where $a=|\bm a|$ and $g_{\parallel} = N_F {\rm Re}[\sinh \alpha \sinh \theta (m_y \cos \delta + m_z \sin \delta)]$. $N_{\hat h_\parallel} = N/2 + g_\parallel$ is the spin-resolved DOS in the direction of $h_\parallel$ (the projection of the exchange field on $(y,z)$-plane). This quantity can be directly extracted from the spatial dependence of the experimentally measured spin-resolved DOS. One can show that the spin-resolved DOS corresponding to the direction $\bm h_\parallel \times \bm e_x$ is zero.

\begin{figure}[!tbh]
    \centerline{\includegraphics[clip=true,width=3.2in]{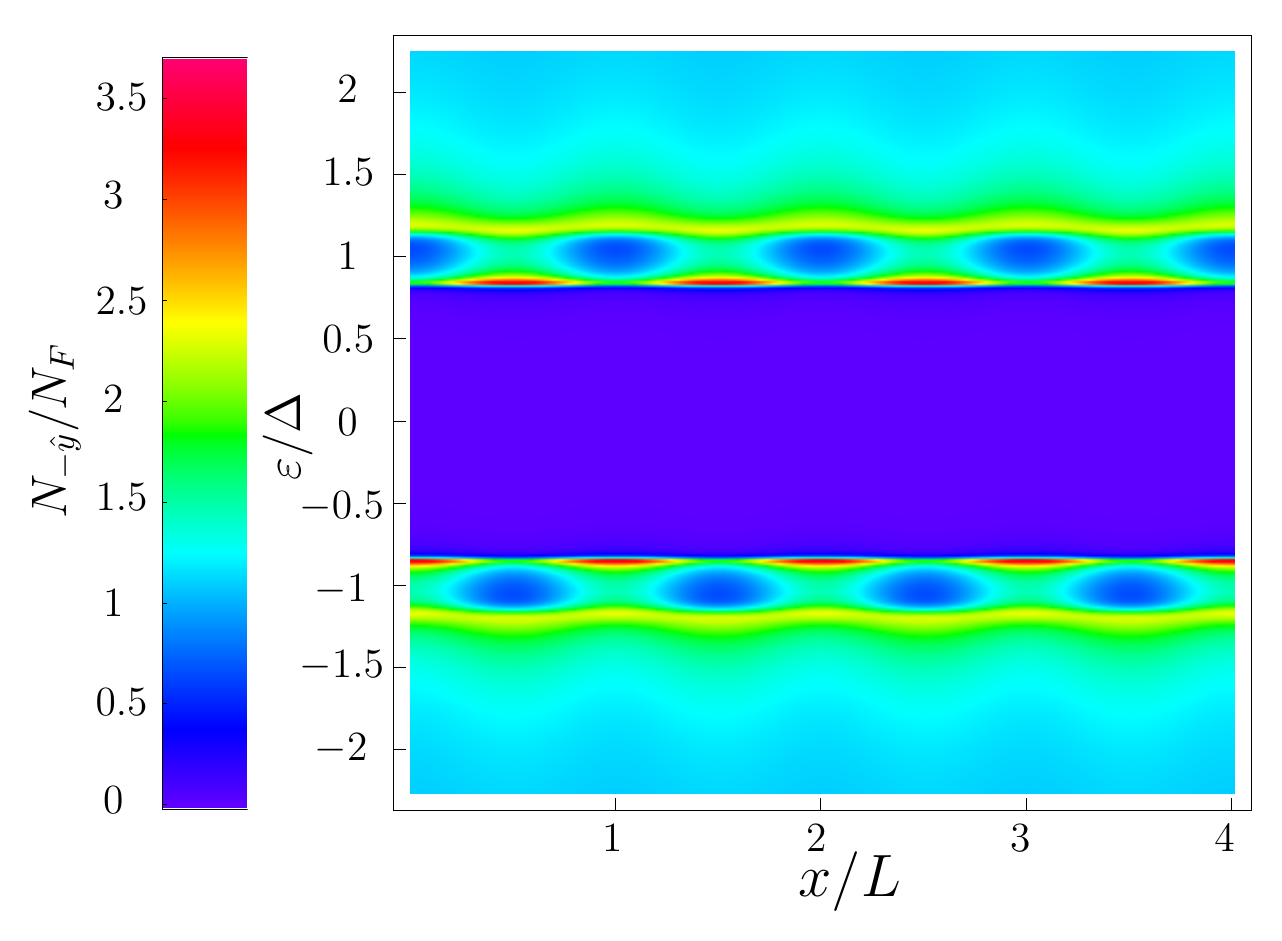}}
      \caption{ 2D plot of the spin-resolved DOS $N_{-\hat y}$ in $(\varepsilon,x)$-plane. $N_{\hat z}$ has the same form with a spatial phase shift $\pi/2$. $h_\textrm{eff} = 0.2 \Delta$, $L/\xi_S = 5$, $\delta = 0.25 \pi$.}
 \label{spatial_helix}
 \end{figure}

We observe that the spin-resolved DOS components $N_{\pm \hat x}$ and $N_{\hat h_\parallel}$ are spatially constant, while $N_{\pm \hat y}$ and $N_{\pm \hat z}$ are spatially modulated with the period of the helix. Therefore, the magnetic tip STM measurements of the DOS in S/F bilayers can give information about the exact spatial structure of the texture. Fig.~\ref{spatial_helix} demonstrates a 2D plot  of the spin-resolved DOS ($N_{\pm \hat y}$ and $N_{\pm \hat z}$ only differ by the spatial shift along the $x$-axis) in $(\varepsilon,x)$-plane for the S/F bilayer with a helix spin texture. The spatial modulations of the inner DOS features with the helix period are clearly seen on the sides of the central featureless gap region. At the same time the outer peaks practically do not depend on the coordinate.

\begin{figure}[!tbh]
 \begin{minipage}[b]{\linewidth}
   \centerline{\includegraphics[clip=true,width=2.8in]{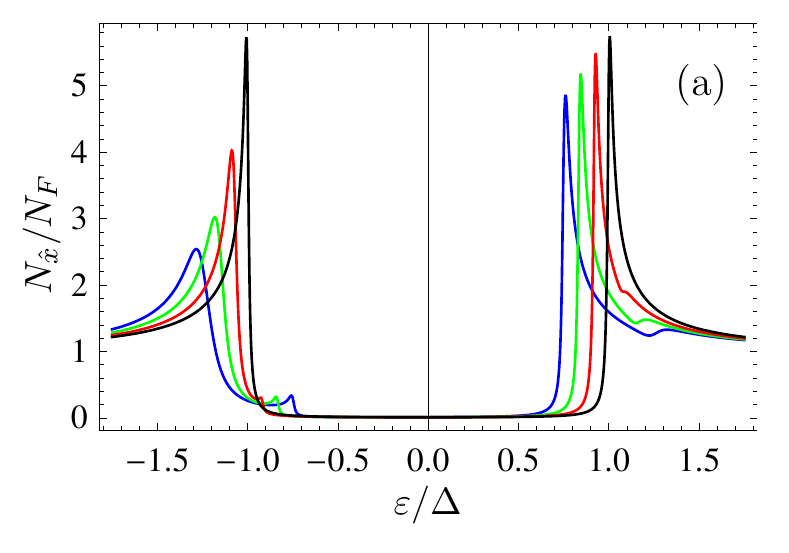}}
   \end{minipage}
   \begin{minipage}[b]{\linewidth}
   \centerline{\includegraphics[clip=true,width=2.8in]{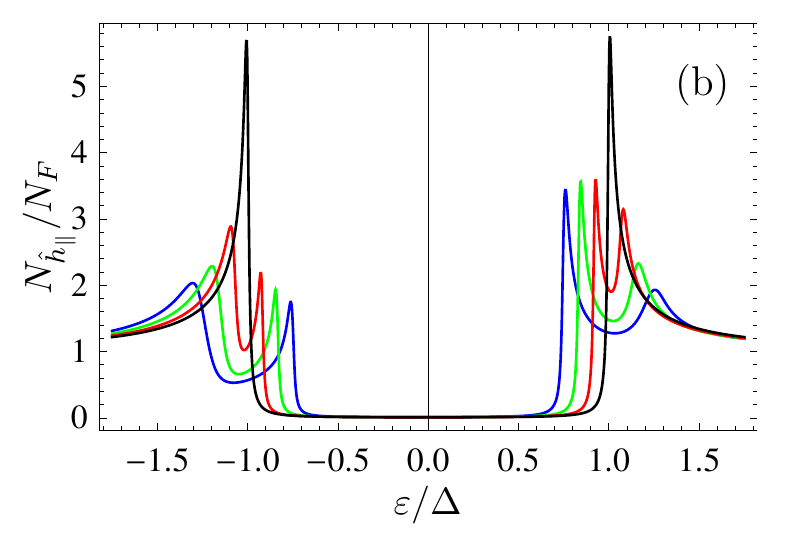}}
   \end{minipage}
   \caption{ Spin-resolved DOS $N_{\hat x}$ [(a)] and spin-resolved DOS $N_{\hat h_\parallel}$ [(b)] as functions of energy for different exchange fields: 0 (black), $0.1 \Delta $ (red), $0.2 \Delta $ (green), $0.3 \Delta $ (blue). $\delta = 0.25 \pi$ and $L/\xi_S = 5$.}
 \label{DOS_spin_h}
 \end{figure}
 
Further on, we concentrate on the dependence of $N_{\hat x}$ and $N_{\hat h_\parallel}$ on the effective exchange field and effective SO coupling. The dependence of $N_{\hat h_\parallel}$ and $N_{\hat x}$ on quasiparticle energy for different $h_\textrm{eff}$ is presented in Fig.~\ref{DOS_spin_h}. We see that they are asymmetric with respect to $\varepsilon \to -\varepsilon$. This situation is typical for the spin-resolved DOS and $N_{\hat h_\parallel}(\varepsilon) = N_{-\hat h_\parallel}(-\varepsilon)$, that is the particle-hole symmetry is restored if one considers the DOS for the opposite spin directions simultaneously. The same is valid for $N_{\hat x}$. The splitting between the coherence peaks in $N_{\hat x}$ and $N_{\hat h_\parallel}$ is the same as for the usual DOS $N$.

\begin{figure}[!tbh]
 \begin{minipage}[b]{\linewidth}
   \centerline{\includegraphics[clip=true,width=2.8in]{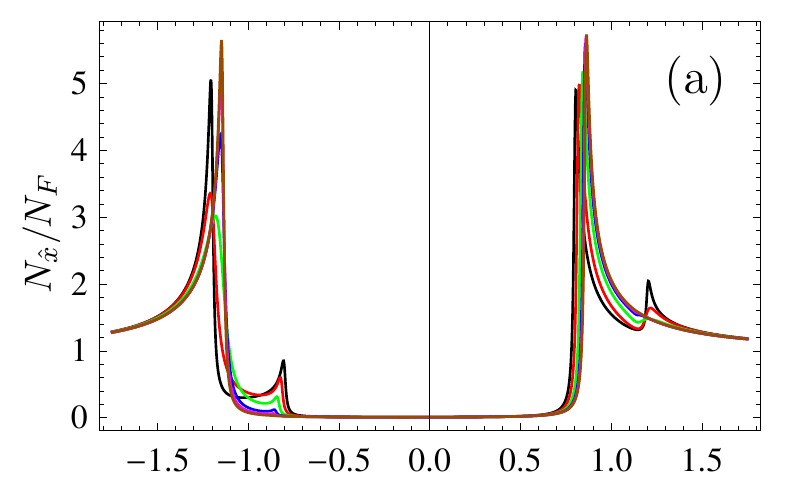}}
   \end{minipage}
   \begin{minipage}[b]{\linewidth}
   \centerline{\includegraphics[clip=true,width=2.8in]{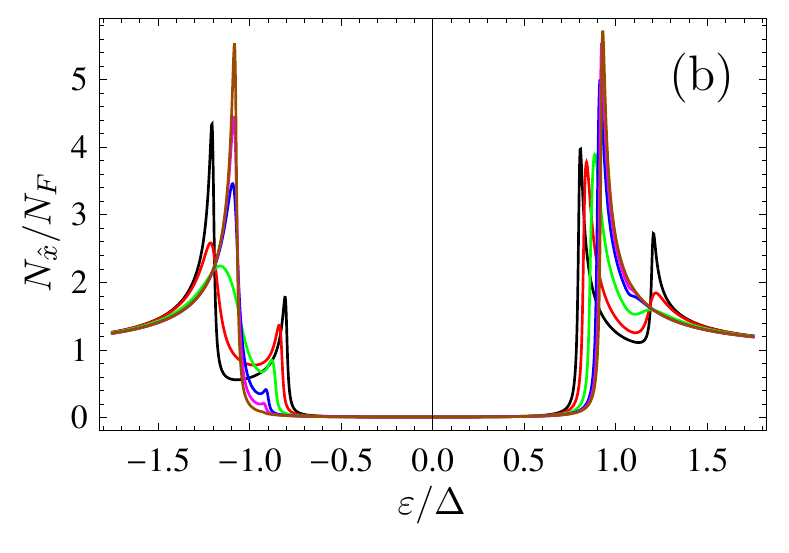}}
   \end{minipage}
   \caption{Spin-resolved DOS $N_{\hat x}$ as a function of energy for different effective SO couplings. (a) $\delta = 0.25 \pi$; (b) $\delta = 0.375 \pi$. $h_\textrm{eff} = 0.2 \Delta$ for the both panels. The color coding is the same as in Fig.~\ref{DOS_SO}.}
 \label{DOS_spin_SO1}
 \end{figure}

\begin{figure}[!tbh]
 \begin{minipage}[b]{\linewidth}
   \centerline{\includegraphics[clip=true,width=2.8in]{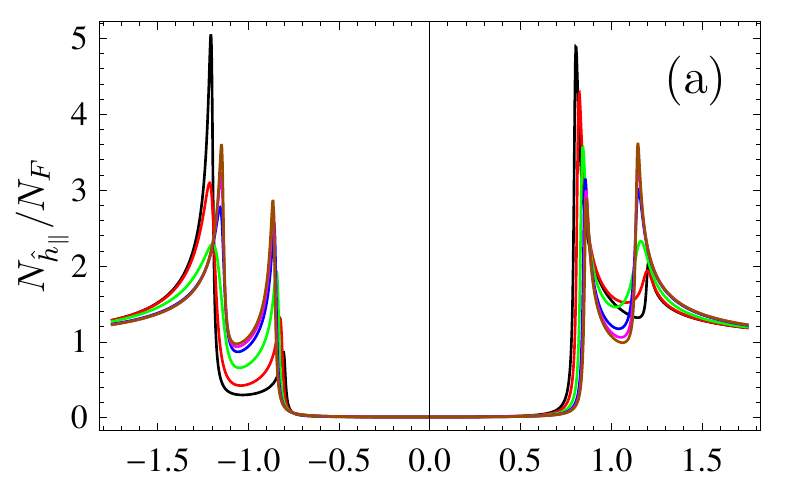}}
   \end{minipage}
   \begin{minipage}[b]{\linewidth}
   \centerline{\includegraphics[clip=true,width=2.9in]{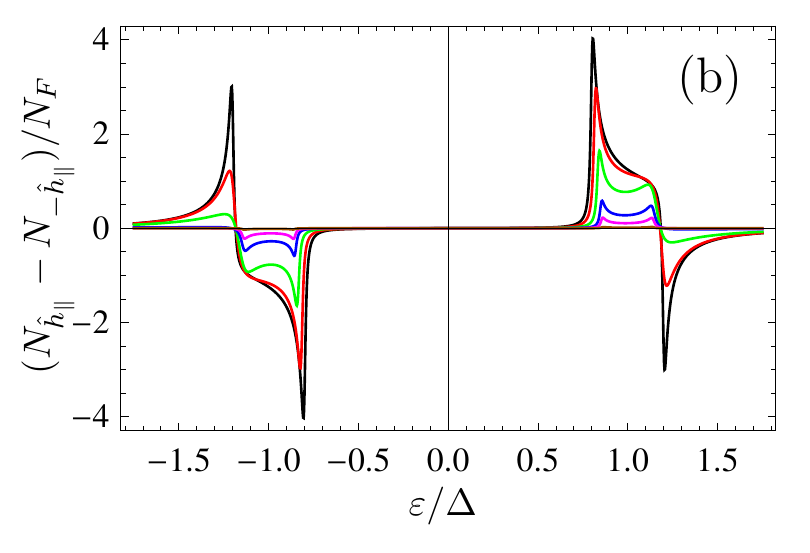}}
   \end{minipage}
   \caption{(a) Spin-resolved DOS $N_{\hat h_\parallel}$ as a function of energy for different effective SO couplings. (b) $N_{\hat h_\parallel} - N_{-\hat h_\parallel}$ as a function of energy for different effective SO couplings. $\delta = 0.25 \pi$, $h_\textrm{eff} = 0.2 \Delta$ for the both panels. The color coding is the same as in Fig.~\ref{DOS_SO}.}
 \label{DOS_spin_SO2}
 \end{figure}

The dependences of $N_{\hat x}$ and $N_{\hat h_\parallel}$ on energy for different values of the effective SO coupling are presented in Figs.~\ref{DOS_spin_SO1} and \ref{DOS_spin_SO2}, respectively. From Fig.~\ref{DOS_spin_SO1} it is seen that upon increasing the effective SO (that is, reducing $L$) the behavior of $N_{\hat x}$ evolves from the typical spin-resolved DOS behavior for the homogeneous superconductor with $\bm h_\textrm{eff} \nparallel \bm e_x$ with split DOS peaks of unequal heights (at very large $L$) to the typical spin-resolved DOS behavior for $\bm h_\textrm{eff} = h_x \bm e_x$, where the peaks are not split, but shifted along the energy axis by the value $h_x$ (at $L \ll \xi_S$).

Fig.~\ref{DOS_spin_SO2} demonstrates the behavior of $N_{\hat h_\parallel}$ on energy for different values of the effective SO coupling. It is seen that upon increasing the effective SO its behavior evolves from asymmetric with respect to the quasiclassical energy to the symmetric one. It is more clearly seen on panel (b) of this figure, where the difference $N_{\hat h_\parallel} - N_{-\hat h_\parallel}$ is presented. This difference goes to zero at $L \ll \xi_S$, what is equivalent to the fact that $N_{\hat h_\parallel}$ becomes symmetric. The natural explanation is that at $L \ll \xi_S$ the component of $h_\textrm{eff}$ located in the $(y,z)$-plane is effectively averaged and the superconductor does not feel it. Therefore, any direction in the $(y,z)$-plane is perpendicular to the effective exchange $h_x \bm e_x$ and we have spin-degenerate (and, consequently, symmetric with respect to $\varepsilon \to -\varepsilon$) DOS for it. Summarizing this part we can conclude that at $L \ll \xi_S$ all the components of the DOS behave as for the superconductor with an effective exchange field directed along the $x$-axis.

\section{DOS in ballistic spin-textured S/F bilayers}

\subsection{Model and method}

The model system is the same as in Fig.~\ref{sketch}. It consists of the spin-textured ferromagnet with the spatially dependent exchange field $\bm h(\bm r)$ and a spin-singlet superconductor. The magnetic proximity effect is again described by adding the effective exchange field $h_\textrm{eff}(\bm r)$ to the quasiclassical equation, which is used for treating the superconductor. The only difference is that here we consider the ballistic case and the S film is described by the Eilenberger equation for the retarded Green's function:
\begin{eqnarray}
 i \bm v_F \nabla\check g(\bm r, \bm p_F)+\Bigl[ \varepsilon \tau_z  + \bm h_\textrm{eff}(\bm r) \bm \sigma \tau_z - \check \Delta,\check g \Bigr] = 0,~~~~~~
 \label{eilenberger}
\end{eqnarray}
Contrary to the problem of magnetic helix texture,  the spin gauge transform in order to reduce the problem to the homogeneous one does not help. Hence, for textures of domain wall type such a gauge transform does not lead to such a simplification of the problem. Therefore, we do not perform the spin gauge transform and solve the problem numerically in the fixed spin basis.

In the ballistic case, it is convenient to use the so-called Riccati parametrization for the Green's function \cite{Eschrig2000,Eschrig2009}. In terms of the Riccati parametrization the retarded Green's function takes the form:
\begin{eqnarray}
\check g =
\left(
\begin{array}{cc}
1+\hat \gamma \hat {\tilde \gamma} & 0 \\
0 & 1+\hat {\tilde \gamma} \hat \gamma \\
\end{array}
\right)^{-1}
\left(
\begin{array}{cc}
1-\hat \gamma \hat {\tilde \gamma} & 2 \hat \gamma \\
2 \hat {\tilde \gamma} & -(1-\hat {\tilde \gamma} \hat \gamma) \\
\end{array}
\right)
\label{riccati_GF}
\end{eqnarray}
where $\hat \gamma$ and $\hat {\tilde \gamma}$ are matrices in spin space. Note that, our parametrization differs from the definition in the literature \cite{Eschrig2000,Eschrig2009} by factors $i\sigma_y$ as $\hat \gamma_{standard} = \hat \gamma i \sigma_y$ and $\hat {\tilde \gamma}_{standard} = i \sigma_y \hat {\tilde \gamma}$.
The Riccati parametrization Eq.~(\ref{riccati_GF}) obeys the normalization condition $\check g^2 = 1$ automatically.

The Riccati amplitudes $\hat \gamma$ and $\hat {\tilde \gamma}$ obey the following Riccati-type equations:
\begin{eqnarray}
 i \bm v_F \nabla \hat \gamma + 2 \varepsilon  \gamma = -\Delta^* \hat \gamma^2 - \bigl\{ \bm h \bm \sigma, \hat \gamma \bigr\} - \Delta ,
 \label{riccati}
\end{eqnarray}
and $\hat {\tilde \gamma}$ obeys the same equation with the substitution $\varepsilon \to -\varepsilon$, $\bm h \to -\bm h$ and $\Delta \to \Delta^*$. In this work, we assume $\Delta = \Delta^*$. Moreover, we neglect the spatial variations of the order parameter and assume $\Delta = const$.

If we consider a locally spatially inhomogeneous  magnetic texture like a domain wall (see Fig.~\ref{sketch}(b)) the Riccati amplitudes $\hat \gamma$ and $\hat {\tilde \gamma}$ can be found from Eq.~(\ref{riccati}) numerically with the following asymptotic condition:
\begin{align}
 \hat \gamma_{\infty}  = & \gamma_{0\infty} + \frac{\bm h_{\infty} \bm \sigma}{h} \gamma_\infty , \\
 \gamma_{0\infty}  = & -\frac{1}{2}\Bigl[ \frac{\Delta}{\varepsilon +h_\textrm{eff}+i\sqrt{\Delta^2 - (\varepsilon + h_\textrm{eff})^2}} \nonumber \\ 
 & + \frac{\Delta}{\varepsilon-h_\textrm{eff}+i\sqrt{\Delta^2 - (\varepsilon  - h_\textrm{eff})^2}} \Bigr],
 \\
 \gamma_\infty = & -\frac{1}{2}\Bigl[ \frac{\Delta}{\varepsilon+h_\textrm{eff}+i\sqrt{\Delta^2 - (\varepsilon  + h_\textrm{eff})^2}} \nonumber \\ & - \frac{\Delta}{\varepsilon -h_\textrm{eff}+i\sqrt{\Delta^2 - (\varepsilon  - h_\textrm{eff})^2}} \Bigr],
  \label{riccati_asymptotic}
\end{align}
and $\hat {\tilde \gamma}_\infty = - \hat \gamma_\infty$.

Eq.~(\ref{riccati}) is numerically stable if it is solved starting from $x = -\infty$ for right-going trajectories $v_x > 0$ and from $x = +\infty$ for left-going trajectories $v_x < 0$. On the contrary, $\hat {\tilde \gamma}$ can be found numerically starting from $x = +\infty$ for right-going trajectories $v_x > 0$ and from $x = -\infty$ for left-going trajectories $v_x < 0$.

Having at hand $\hat \gamma(x,\bm p_F)$ and $\hat {\tilde \gamma}(x, \bm p_F)$ it is possible to calculate the DOS and spin resolved DOS (normalized to the normal state DOS) as follows:
\begin{eqnarray}
N= N_F{\rm Re}\Bigl\{{\rm Tr}\langle \hat g \rangle\Bigr\}
\label{DOS_ballistic}
\end{eqnarray}
and the spin-resolved DOS corresponding to the spin along the direction $\bm l$ is calculated as
\begin{eqnarray}
N_{\hat l}=\frac{N_F}{2}{\rm Re}\Bigl\{ {\rm Tr}\langle \hat g \rangle +
\bm l_i {\rm Tr}\langle \sigma_i \hat g \rangle \Bigr\},
\label{DOS_spin_resolved_ballistic}
\end{eqnarray}
where $\langle ... \rangle$ means averaging over the 2D Fermi surface of the superconducting film and $\hat g = (1+ \hat \gamma \hat {\tilde \gamma})^{-1}(1 - \hat \gamma \hat {\tilde \gamma})$ is the normal Green's function, that is the upper left element in the particle-hole space of Eq.~(\ref{riccati_GF}).

\subsection{Results: domain wall}

It is convenient to parametrize the spin texture of the head-to-head (or tail-to-tail) DW by
\begin{eqnarray}
\bm h = h (\cos \theta, \sin \theta \cos \delta, \sin \theta \sin \delta),
\label{h_texture_hhwall}
\end{eqnarray}
where in general the both angles depend on $x$-coordinate. The equilibrium shape of the DW is dictated by the interplay between the magnetic anisotropy energy and the exchange energy and is given by
 \begin{equation} \label{theta}
 \cos\theta = \pm  \tanh [x/d_w],
 \end{equation}
and $\delta = const$ for the classical in-plane DW. However, here we also consider noncoplanar DW configurations, which correspond to a spatially dependent $\delta$. This situation can occur due to the contact with the superconductor or due to external perturbations such as applied supercurrent or applied magnetic field.

\begin{figure}[!tbh]
 \begin{minipage}[b]{\linewidth}
   \centerline{\includegraphics[clip=true,width=2.5in]{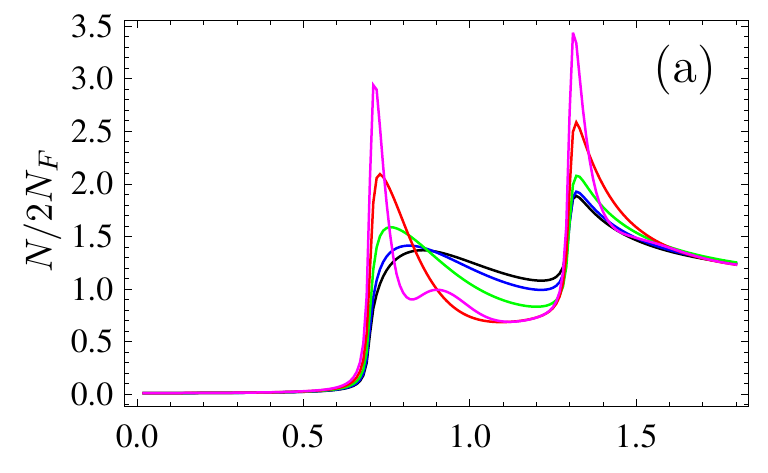}}
   \end{minipage}
   \begin{minipage}[b]{\linewidth}
   \centerline{\includegraphics[clip=true,width=2.5in]{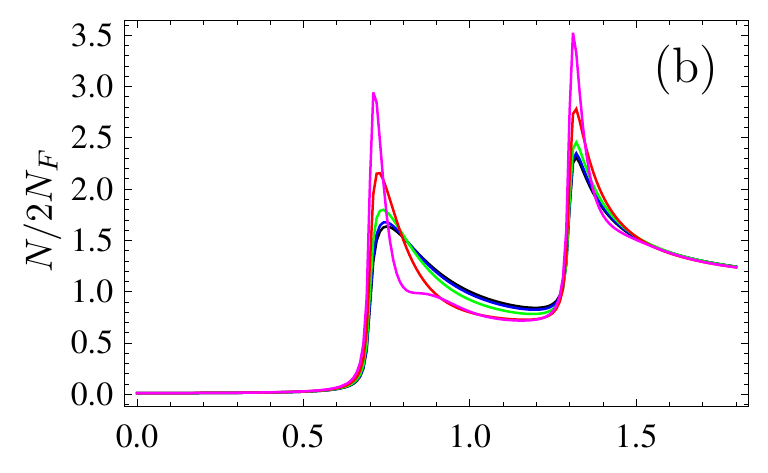}}
   \end{minipage}
   \begin{minipage}[b]{\linewidth}
   \centerline{\includegraphics[clip=true,width=2.5in]{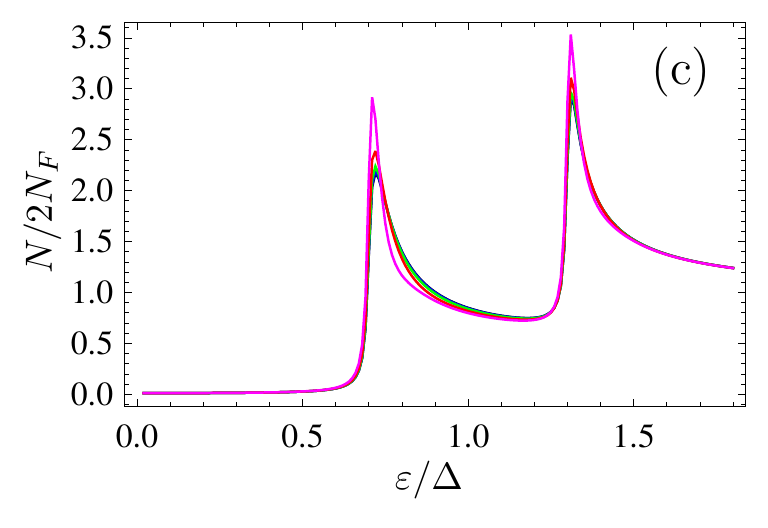}}
   \end{minipage}
   \caption{DOS for a noncoplanar domain wall with $\delta(x) = \pi/(4\cosh^2[x/2d_w]) $ as a function of quasiparticle energy. Only $\varepsilon >0$ data are plotted, while negative energies follow from symmetry. Different curves correspond to different distances $x$ from the wall centre: $x=0$ (black), $x=0.5 \xi_S$ (blue), $x=\xi_S$ (green), $x=2 \xi_S$ (red), $x=5 \xi_S$ (pink).] (a) $d_w = 0.5 \xi_S$; (b) $d_w = \xi_S$; (c) $d_w = 2 \xi_S$. $h_\textrm{eff}=0.3 \Delta$ for all the panels.}
 \label{DOS_DW}
 \end{figure}

The DOS for a weakly noncoplanar wall is presented in Fig.~\ref{DOS_DW}. Different curves correspond to different distances from the wall centre. The results for a coplanar wall with $\delta = 0$ are practically the same and are not presented here. Obviously, far from the wall centre the DOS is very close to the double-split DOS of a homogeneous superconductor with $\bm h = \bm h_\textrm{eff}$. The exception is the Friedel-type oscillation \cite{Tomasch1965,Rowell1966}, which decays very slowly in the energy region close to the gap and are a signature of the ballistic system. The DOS for wide enough DWs (see Fig.~\ref{DOS_DW}(c), where $d_w = 2 \xi_S$) is close to the shape typical for a homogeneous superconductor with the exchange field coinciding with the local field at a given point. That is, if the characteristic width of the wall is larger than $(1 \div 2) \xi_S$, the spin-independent DOS a superconductor results from an effectively homogeneous exchange field. However, this picture is not correct for the spin-resolved DOS, what is discussed below.  

\begin{widetext}

\begin{figure}[!tbh]
 \includegraphics[clip=true,width=6.6in]{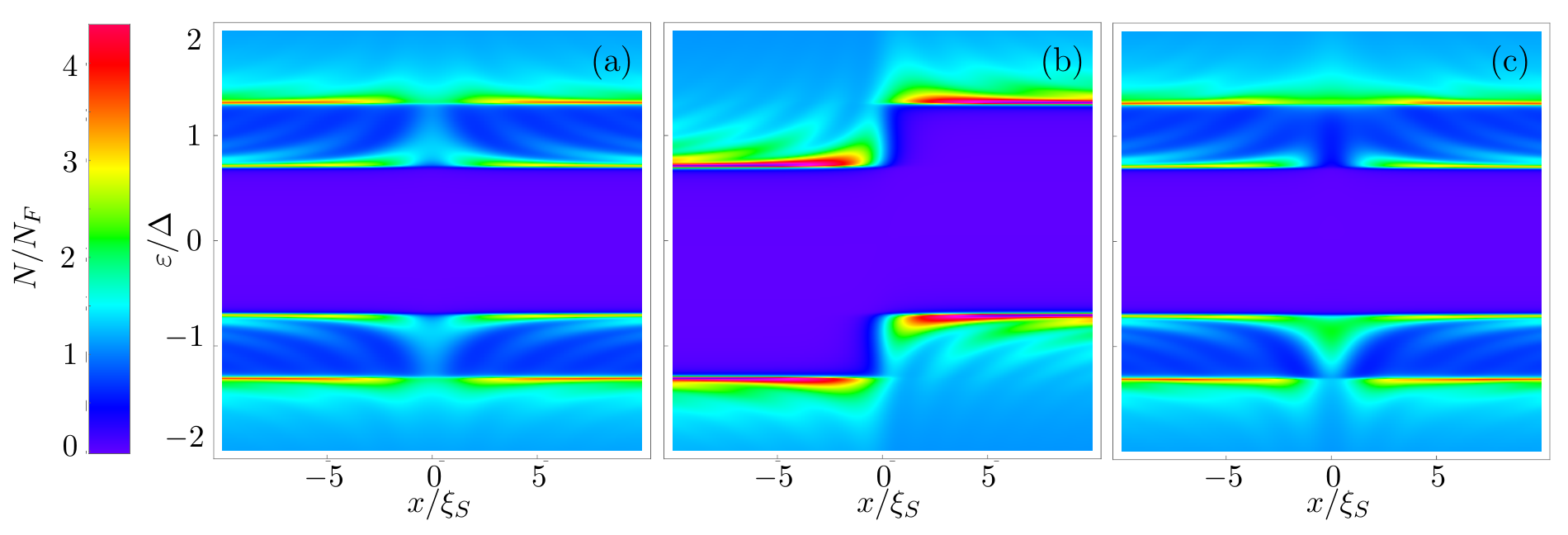}
   \caption{2D plot of (a) DOS $N$; (b) spin-resolved DOS $N_{-\hat x}$  and (c) $N_{-\hat y}$ in $(\varepsilon,x)$-plane. $h_\textrm{eff} = 0.3 \Delta$, $d_W/\xi_S = 0.5$, The noncoplanarity $\delta(x)$ is the same as in Fig.~\ref{DOS_DW}.}
 \label{2D_2}
 \end{figure}
 
\end{widetext}

Only for the wall widths $d_W \lesssim \xi_S$ the Zeeman-split shape is distorted and we see signatures of the mixing between the peaks in the vicinity of the wall. The 2D plot of the DOS for such a short wall with $d_W = 0.5 \xi_S$ is presented in Fig.~\ref{2D_2}(a). The mixing between the coherence peaks is clearly seen in the region occupied by the wall. 

 \begin{figure}[!tbh]
 \begin{minipage}[b]{0.5\linewidth}
   \centerline{\includegraphics[clip=true,width=1.6in]{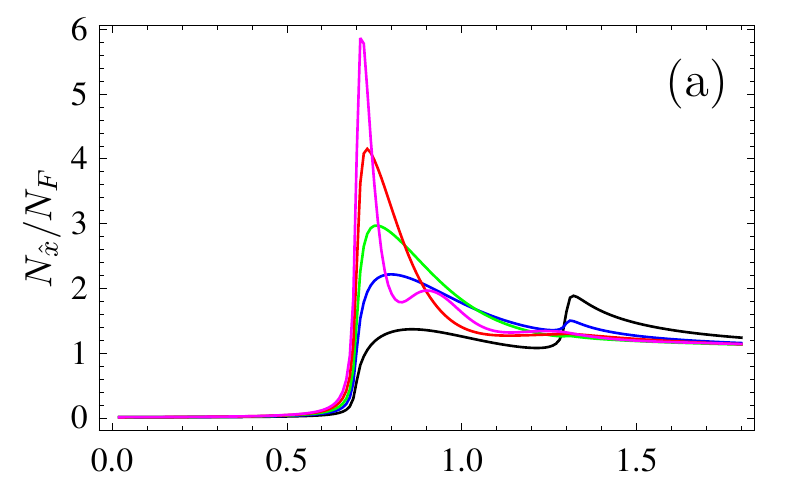}}
   \end{minipage}\hfill
   \begin{minipage}[b]{0.5\linewidth}
   \centerline{\includegraphics[clip=true,width=1.6in]{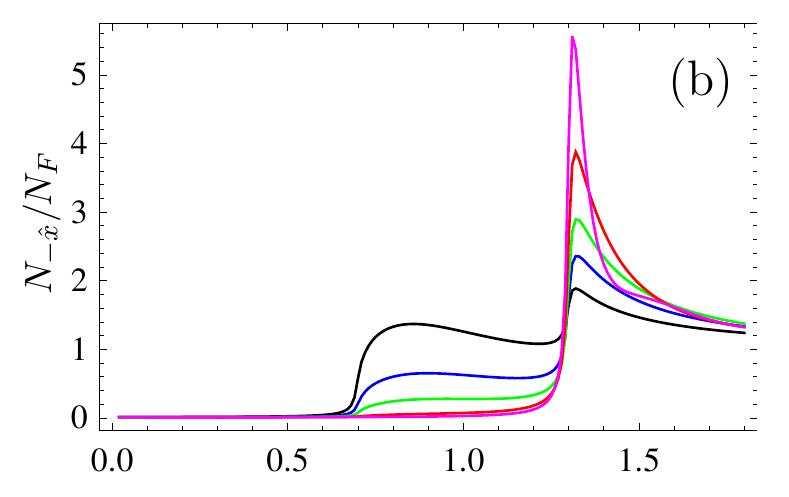}}
   \end{minipage}
   \begin{minipage}[b]{0.5\linewidth}
   \centerline{\includegraphics[clip=true,width=1.6in]{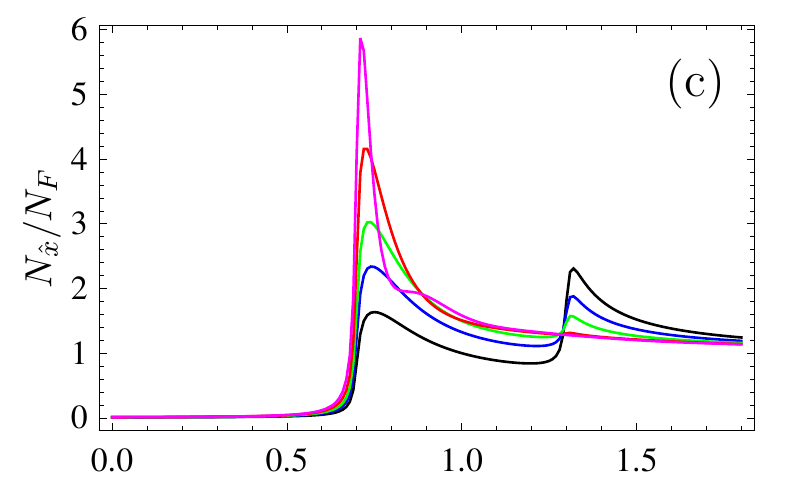}}
   \end{minipage}\hfill
   \begin{minipage}[b]{0.5\linewidth}
   \centerline{\includegraphics[clip=true,width=1.6in]{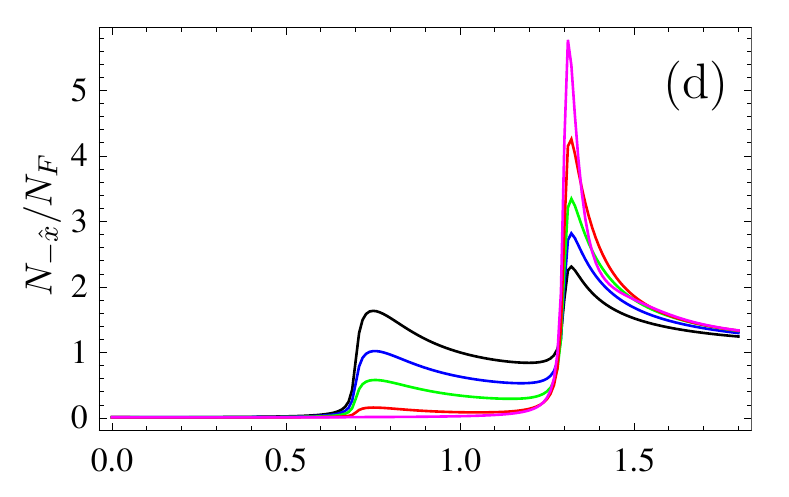}}
   \end{minipage}
   \begin{minipage}[b]{0.5\linewidth}
   \centerline{\includegraphics[clip=true,width=1.6in]{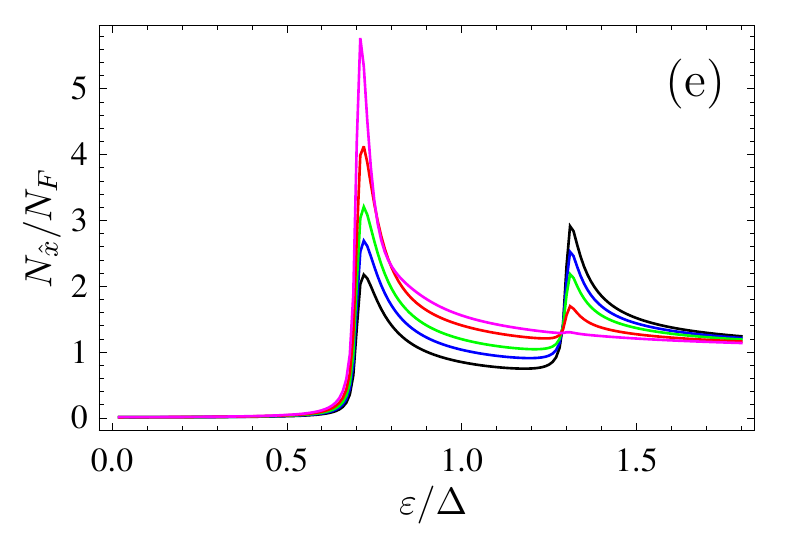}}
   \end{minipage}\hfill
   \begin{minipage}[b]{0.5\linewidth}
   \centerline{\includegraphics[clip=true,width=1.6in]{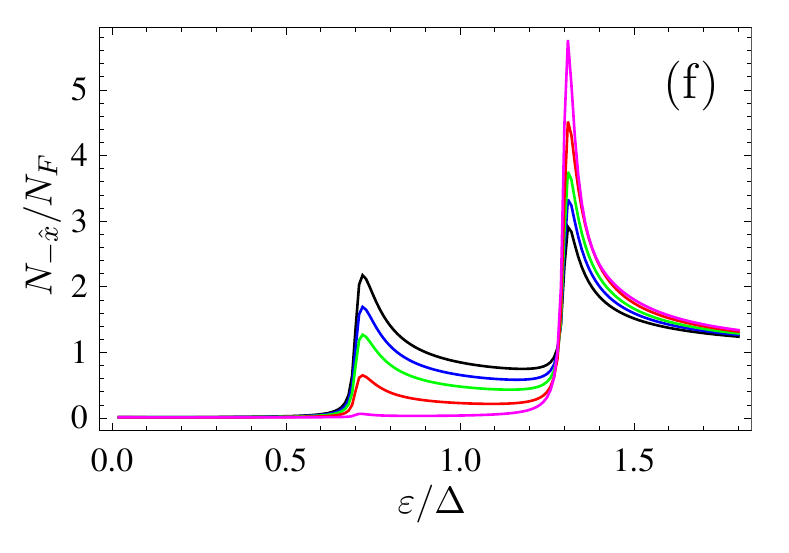}}
   \end{minipage}
   \caption{Spin-resolved DOS $N_{\hat x}$ (left column) and $N_{-\hat x}$ (right column) for noncoplanar domain wall as a function of quasiparticle energy. Only $\varepsilon >0$ data are plotted. $\varepsilon<0$ part of the curve can be obtained by symmetry $N_{\hat x}(-\varepsilon) = N_{-\hat x}(\varepsilon)$. Different curves correspond to different distances $x$ from the wall centre: $x=0$ (black), $x=0.5 \xi_S$ (blue), $x=\xi_S$ (green), $x=2 \xi_S$ (red), $x=5 \xi_S$ (pink). (a)-(b) $d_w = 0.5 \xi_S$; (c)-(d) $d_w = \xi_S$; (e)-(f) $d_w = 2 \xi_S$. $h_\textrm{eff}=0.3 \Delta$ for all the panels.}
 \label{DOS_spin_x_DW}
 \end{figure}

 The spin-resolved DOS projected on the  $x$-axis, coinciding with the direction of the magnetization in the bulk of the domains, is presented in Fig.~\ref{DOS_spin_x_DW}. The results for the coplanar wall are practically the same and are not presented. We see again that far from the wall centre (pink curves) the DOS is practically the same as the spin-resolved DOS for a homogeneous superconductor with $\bm h = h_\textrm{eff} \bm e_x$. The results for a narrow DW $d_w = 0.5 \xi_S$ are qualitatively similar to the results, obtained in Ref.~\onlinecite{Strambini2017} for a infinitely narrow DW. In contrast to the spin-independent DOS, here the spin-resolved DOS does not follow the homogeneous shape of the local field even for the case of wide walls. The profile of the spin-resolved DOS is always distorted in the vicinity of the centre regardless of the wall width.

 \begin{figure}[!tbh]
 \begin{minipage}[b]{0.5\linewidth}
   \centerline{\includegraphics[clip=true,width=1.6in]{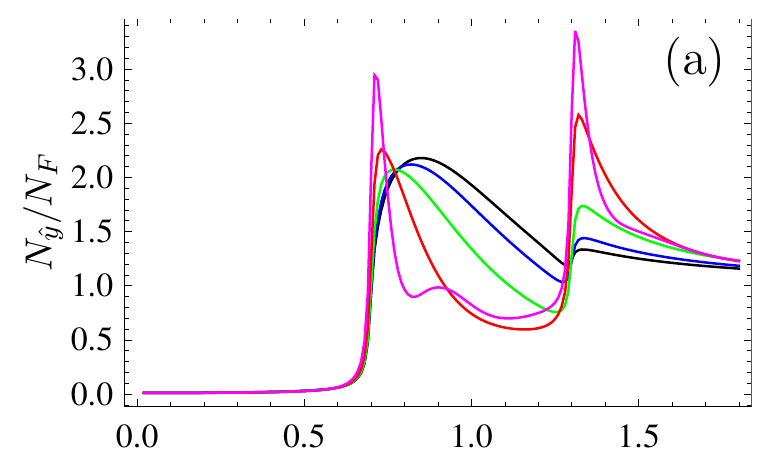}}
   \end{minipage}\hfill
   \begin{minipage}[b]{0.5\linewidth}
   \centerline{\includegraphics[clip=true,width=1.6in]{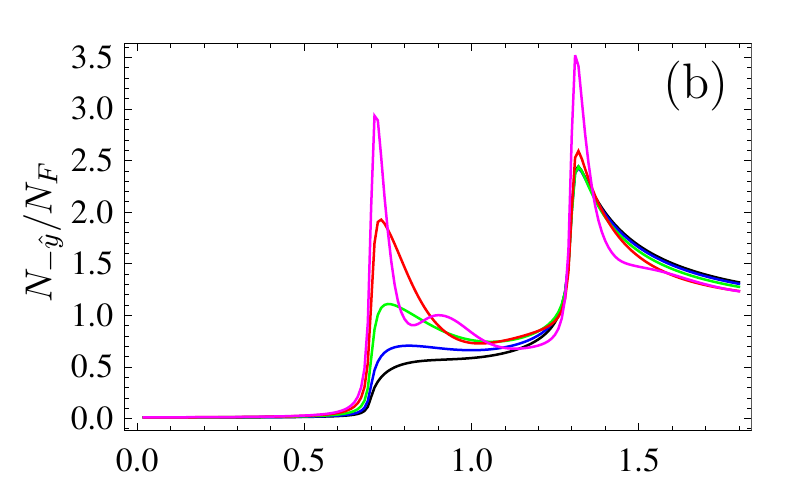}}
   \end{minipage}
   \begin{minipage}[b]{0.5\linewidth}
   \centerline{\includegraphics[clip=true,width=1.6in]{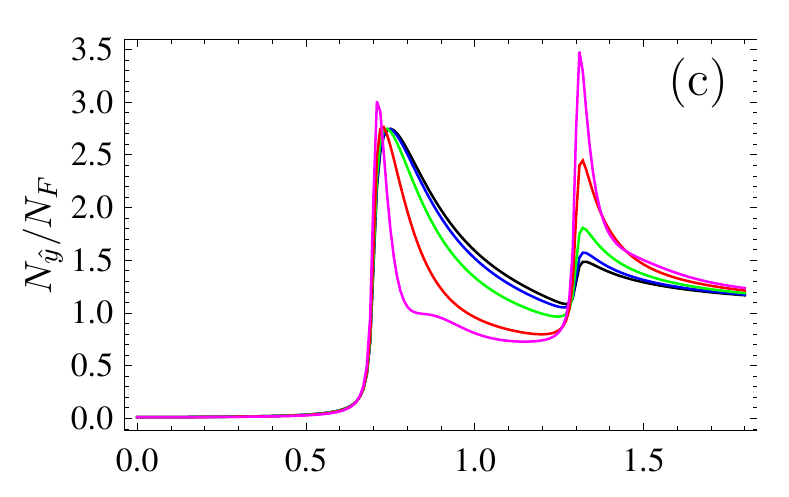}}
   \end{minipage}\hfill
   \begin{minipage}[b]{0.5\linewidth}
   \centerline{\includegraphics[clip=true,width=1.6in]{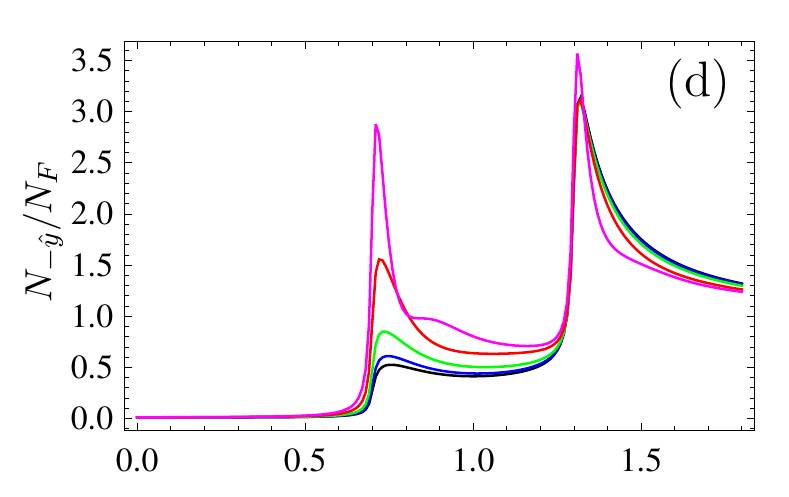}}
   \end{minipage}
   \begin{minipage}[b]{0.5\linewidth}
   \centerline{\includegraphics[clip=true,width=1.6in]{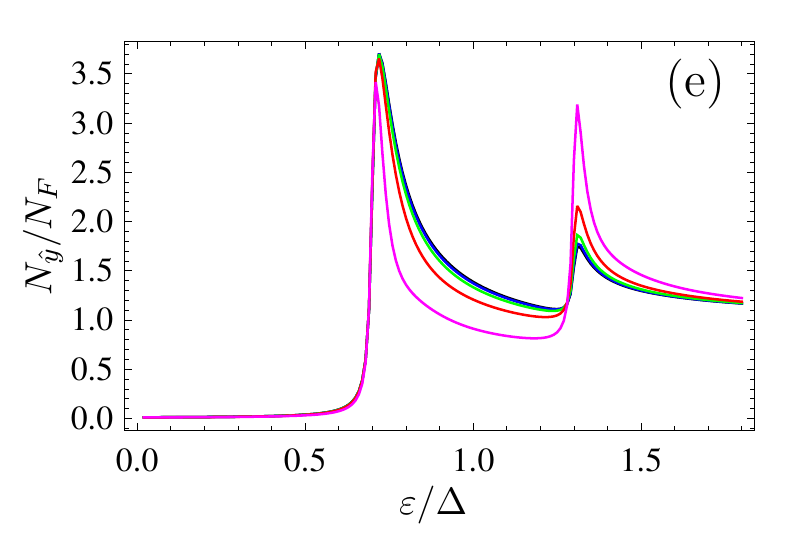}}
   \end{minipage}\hfill
   \begin{minipage}[b]{0.5\linewidth}
   \centerline{\includegraphics[clip=true,width=1.55in]{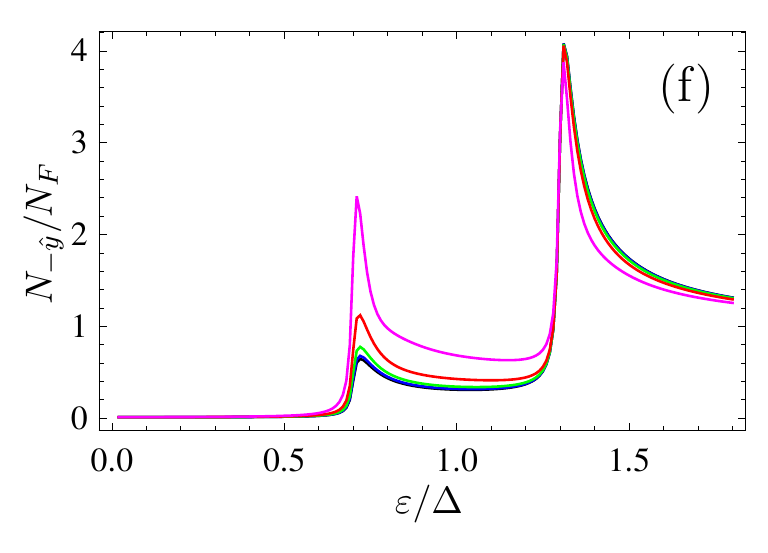}}
   \end{minipage}
   \caption{The spin-resolved DOS $N_{\hat y}$ (left column) and $N_{-\hat y}$ (right column) for parameters similar to  Fig.~\ref{DOS_spin_x_DW}.}
 \label{DOS_spin_y_DW}
 \end{figure}

 The spin-resolved DOS along $y$-axis is presented in Fig.~\ref{DOS_spin_y_DW}. The results for the coplanar wall are practically the same and are not presented. We observe that far from the wall centre (pink curves) the DOS is practically the same as the spin-independent DOS for a homogeneous superconductor. That is because the spin-resolved DOS for a spin direction perpendicular to the direction of the homogeneous exchange field does not depend on spin and here we have a superconductor in a homogeneous exchange field $\bm h = h_\textrm{eff} \bm e_x$ far from the wall. $N_{\pm \hat y}$  also does not follow the homogeneous shape of the local field even for the case of wide walls. The profile of the spin-resolved DOS is always distorted in the vicinity of the centre regardless of the wall width. This component of the spin-resolved DOS cannot be obtained in the model of infinitely narrow wall, because in this model the magnetization is always collinear and there is no spin rotation, which leads to the creation of the equal-spin pairs and, therefore, a nonzero difference $N_{\hat y} - N_{-\hat y}$. The 2D plots of the spin resolved DOS $N_{-\hat x}$ and $N_{-\hat y}$ are presented in Fig.~\ref{2D_2}(b) and (c), respectively.

The spin-resolved DOS along $z$-axis $N_{\hat z}$ and $N_{-\hat z}$ for noncoplanar DW are very similar to $N_{\hat y}$ and $N_{-\hat y}$ and  not presented here. For a coplanar DW with $\delta=0$ (the DW is in $(x,y)$-plane) $N_{\hat z} = N_{-\hat z} = N$ and, therefore, the DOS does not show a spin polarization along the axis perpendicular to the DW plane.
 
 \section{Spin-resolved DOS as a signature of triplet superconducting correlations}

Now our goal is to demonstrate that measurements of the spin-resolved DOS provide an information about the presence and the structure of spin-triplet correlations in the system. Indeed, from the normalization condition
\begin{eqnarray}
\check g^2 =
\left(
\begin{array}{cc}
g_0 + \bm g \bm \sigma & f_0 + \bm f \bm \sigma \\
{\tilde f}_0 + \tilde {\bm f} \bm \sigma & {\tilde g}_0 + \tilde {\bm g} \bm \sigma
\end{array}
\right)^2 = 1,
\label{norm_ph}
\end{eqnarray}
where the structure of the Green's function in particle-hole space is written explicitly, it follows that
\begin{eqnarray}
\bm g = -\frac{1}{2g_0}\bigl[ f_0 \tilde {\bm f} + \tilde f_0 \bm f + i \bm f \times \tilde {\bm f} \bigr].
\label{g_f_connection}
\end{eqnarray}
On the other hand,
\begin{eqnarray}
N_{\hat l}-N_{-\hat l}=N_F {\rm Re} \bigl[ \langle \bm g \rangle \bm l \bigr].
\label{DOS_spin_f_connection}
\end{eqnarray}

Combining Eqs.~(\ref{g_f_connection}) and (\ref{DOS_spin_f_connection}), we can conclude a that nonzero difference $N_{\hat l}-N_{-\hat l}$ is an direct signature of the presence of triplet superconducting correlations in the system. It is worth to note that the inverse statement is not always true: from $N_\uparrow - N_\downarrow = 0$ it does not follow that the odd-frequency correlations are absent, what is similar to the situation in the Zeeman-split conventional superconductors \cite{Linder2015}. This situation is realized in the region of the gap.

{\it Diffusive limit.} In the diffusive limit Eq.~(\ref{DOS_spin_f_connection}) can be further simplified. As it seen from Eq.~(\ref{parametrization}), in this case $\tilde {\bm f} = - \bm f$ and $\tilde f_0 = -f_0$ and the Green's function does not depend on the momentum direction. Hence, we obtain:
\begin{eqnarray}
N_{\hat l}-N_{-\hat l}=N_F {\rm Re} [\tanh \theta \bm f \bm l].
\label{DOS_spin_f_connection_dirty}
\end{eqnarray}
We observe that measurements of spin-resolved DOS corresponding to the spin axis $\bm l$ provide information about the projection of the triplet pair correlations vector $\bm f$ onto the direction $\bm l$.

\begin{widetext}

\begin{figure}[!tbh]
 \begin{minipage}[b]{0.33\linewidth}
   \centerline{\includegraphics[clip=true,width=2.35in]{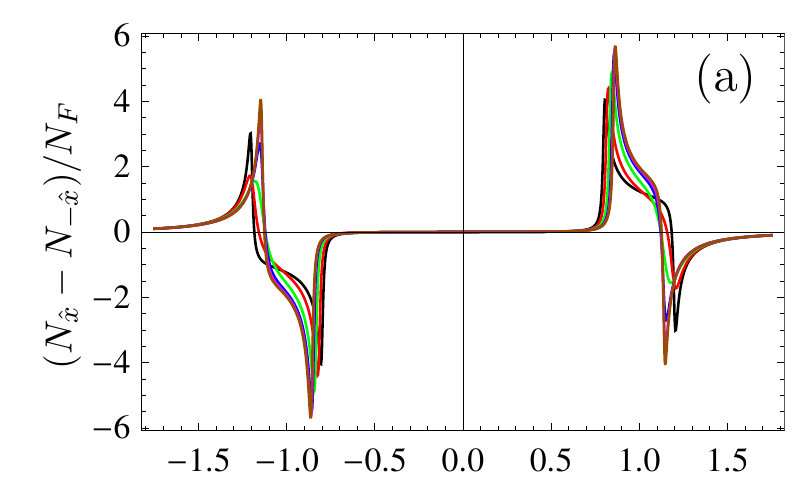}}
   \end{minipage}\hfill
   \begin{minipage}[b]{0.33\linewidth}
   \centerline{\includegraphics[clip=true,width=2.25in]{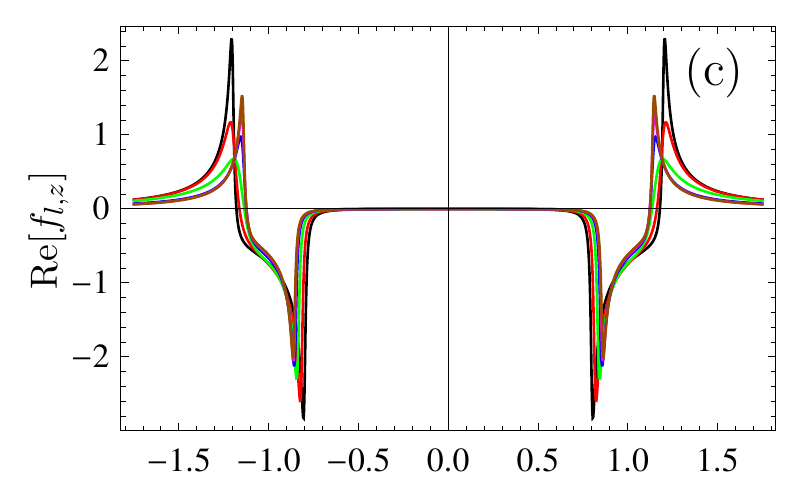}}
   \end{minipage}\hfill
   \begin{minipage}[b]{0.33\linewidth}
   \centerline{\includegraphics[clip=true,width=2.3in]{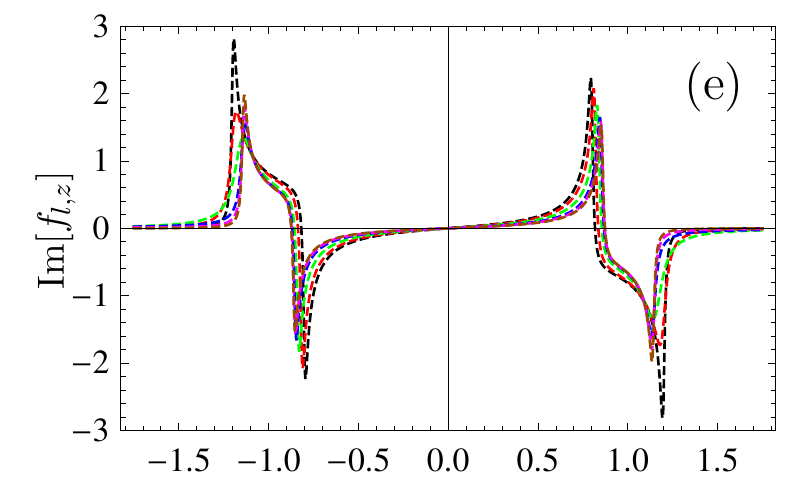}}
   \end{minipage}
   \begin{minipage}[b]{0.33\linewidth}
   \centerline{\includegraphics[clip=true,width=2.25in]{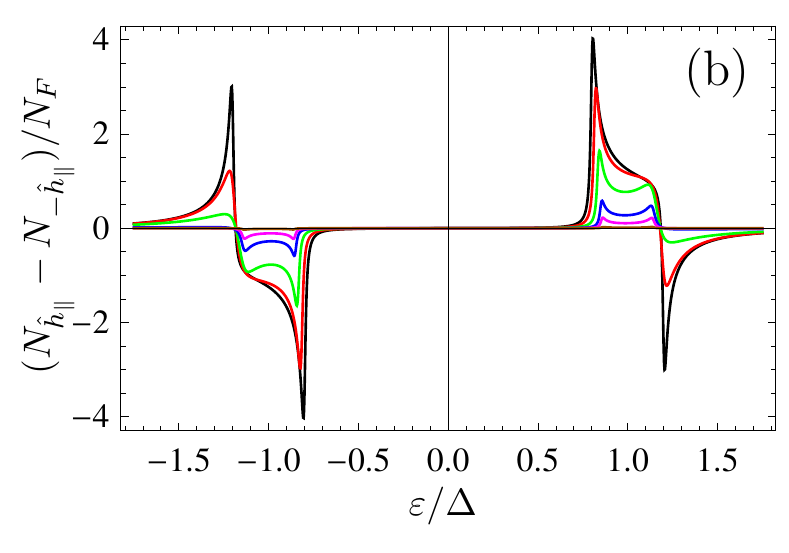}}
   \end{minipage}\hfill
   \begin{minipage}[b]{0.33\linewidth}
   \centerline{\includegraphics[clip=true,width=2.3in]{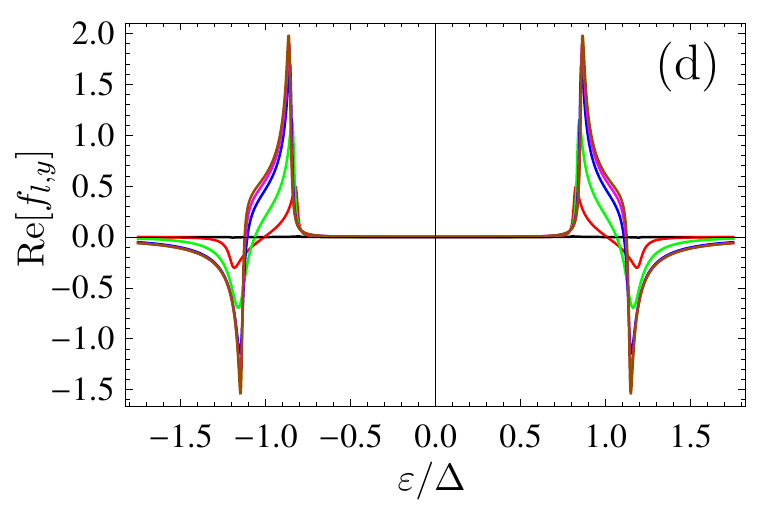}}
   \end{minipage}\hfill
   \begin{minipage}[b]{0.33\linewidth}
   \centerline{\includegraphics[clip=true,width=2.25in]{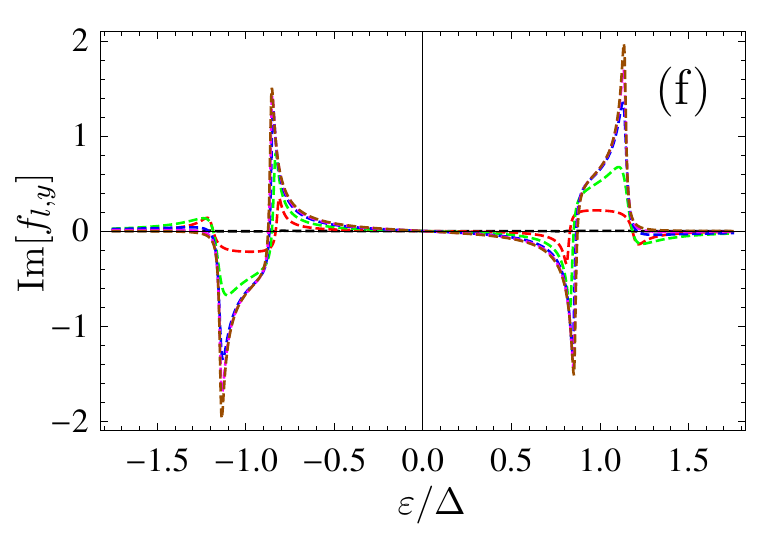}}
   \end{minipage}
   \caption{(a) $N_{\hat x} - N_{-\hat x}$; (b) $N_{\hat h_\parallel} - N_{-\hat h_\parallel}$; (c)-(d) real part of opposite-spin triplet correlations ${\rm Re}[f_{l,z}]$ and equal-spin triplet correlations ${\rm Re}[f_{l,y}]$, respectively; (e)-(f) imaginary part of opposite-spin triplet correlations ${\rm Im}[f_{l,z}]$ and equal-spin triplet correlations ${\rm Im}[f_{l,y}]$, respectively;  Different curves correspond to different values of the effective SO coupling $L/\xi_S = 100$ (black), $10$ (red), $5$ (green), $2.5$ (blue), $1.5$ (pink), $0.5$ (brown). $h_\textrm{eff} = 0.2 \Delta$ and $\delta = 0.25 \pi$ for all the panels.}
 \label{DOS_f_dirty}
 \end{figure}

\end{widetext}

For the case of the magnetic helix, it is physically more reasonable to consider the pair correlations structure in the local spin basis, where the quantization axis coincides with the local magnetization direction. In this basis $\bm f_l \bm \sigma= U^\dagger \bm f \bm \sigma U$ and $f_{l,z} = m_z \sinh \alpha \cosh \theta$ represents the opposite-spin pair correlations $f_{l,z} = (f_{\uparrow \downarrow}+f_{\downarrow \uparrow})/2$, while $f_{l,y} = m_y \sinh \alpha \cosh \theta = -i f_{\uparrow \uparrow} = -i f_{\downarrow \downarrow}$ is the amplitude of the equal-spin pair correlations. Therefore, we can conclude that measurements of the spin-resolved DOS along the direction of the local magnetization $\bm l = \bm h$ provide us with information about the mixed-spin triplet correlations without the contamination of the equal-spin correlations, while measurements of the spin-resolved DOS along $\bm l$ perpendicular to the local magnetization direction can give us the information about the equal-spin correlations without the contamination of the mixed-spin correlations signal. At the same time, in the considered case of the helix spin texture in the fixed reference frame the opposite-spin and equal-spin correlations are represented as an oscillating spatial mixture. Therefore, for a certain fixed $\bm l$ direction a mixture of the mixed- and equal-spin correlations should be observed.

Fig.~\ref{DOS_f_dirty} demonstrates the mixed-spin and equal-spin pairing amplitudes in comparison to the spin-resolved DOS. This figure illustrates our statement that in the energy intervals, where the spin-resolved DOS in non-zero, the triplet pair correlations are inevitably present. Therefore, the spin-resolved DOS definitely indicates the presence of the triplet correlations, but does not allow for obtaining their exact structure.

{\it Ballistic limit.} In the clean case the situation is similar. However, the condition $\bm f \times \tilde {\bm f} = 0$ is not fulfilled for each of the trajectories separately. Therefore, the non-unitary pairing \cite{Tkachov2017} with nonzero spin of the pair is nonzero if we are only interested in the direction-resolved pairing before averaging over trajectories. However, the term $\bm f \times \tilde {\bm f} $ becomes zero after averaging over trajectories. This pairing can manifest itself in the DOS under the applying of the supercurrent to the system, which we do not discuss here. The resulting relationship between the spin-resolved DOS and the triplet correlations takes essentially the same form as in the diffusive case:
\begin{eqnarray}
N_{\hat l}-N_{-\hat l}=- N_F {\rm Re} \bigl[ \langle \frac{\tilde f_0 \bm f}{g_0} \rangle \bigr].
\label{DOS_spin_f_connection_ballistic}
\end{eqnarray}

\begin{widetext}

\begin{figure}[!tbh]
 \begin{minipage}[b]{0.33\linewidth}
   \centerline{\includegraphics[clip=true,width=2.3in]{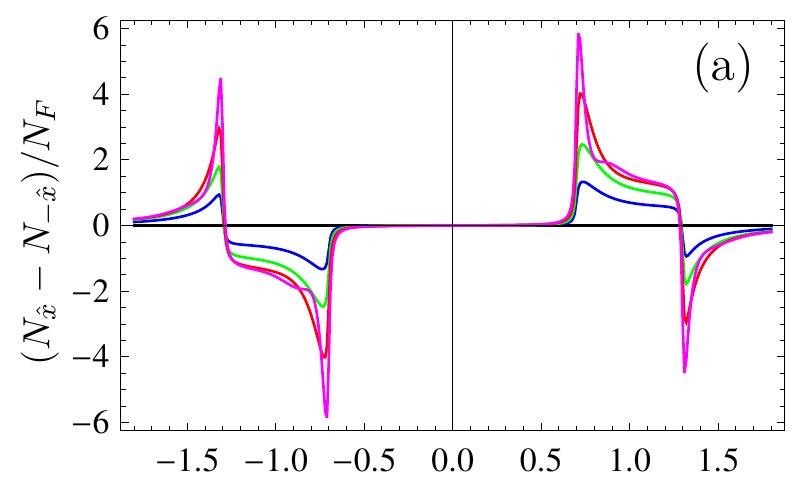}}
   \end{minipage}\hfill
   \begin{minipage}[b]{0.33\linewidth}
   \centerline{\includegraphics[clip=true,width=2.3in]{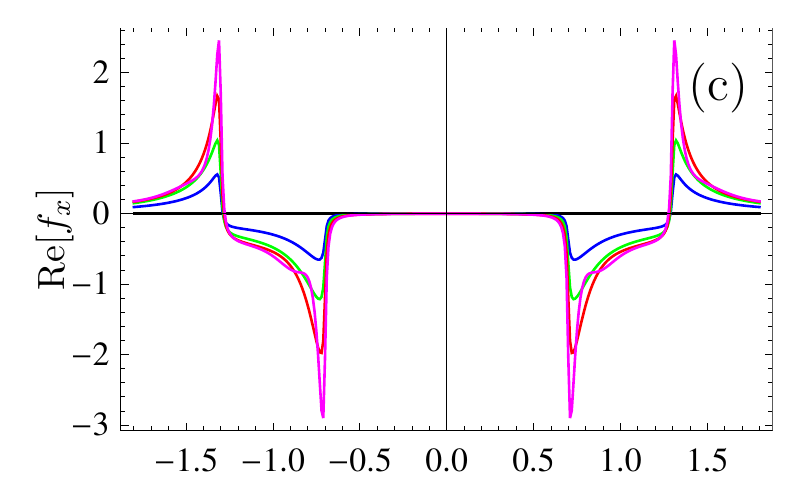}}
   \end{minipage}\hfill
   \begin{minipage}[b]{0.33\linewidth}
   \centerline{\includegraphics[clip=true,width=2.3in]{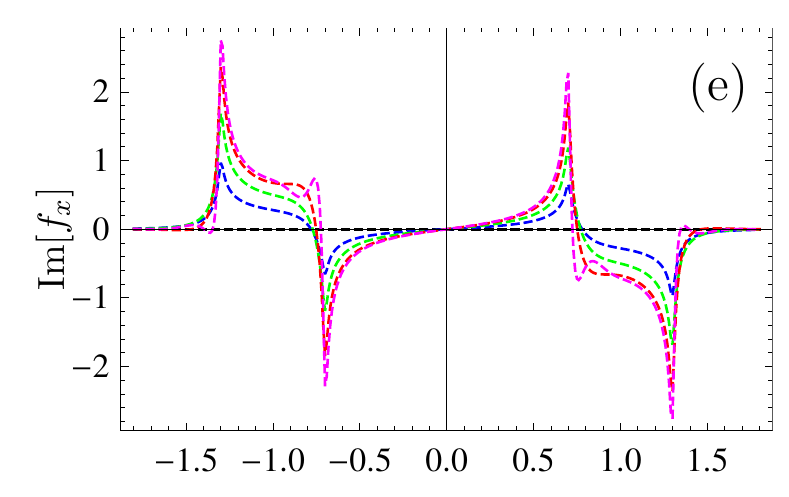}}
   \end{minipage}
   \begin{minipage}[b]{0.33\linewidth}
   \centerline{\includegraphics[clip=true,width=2.3in]{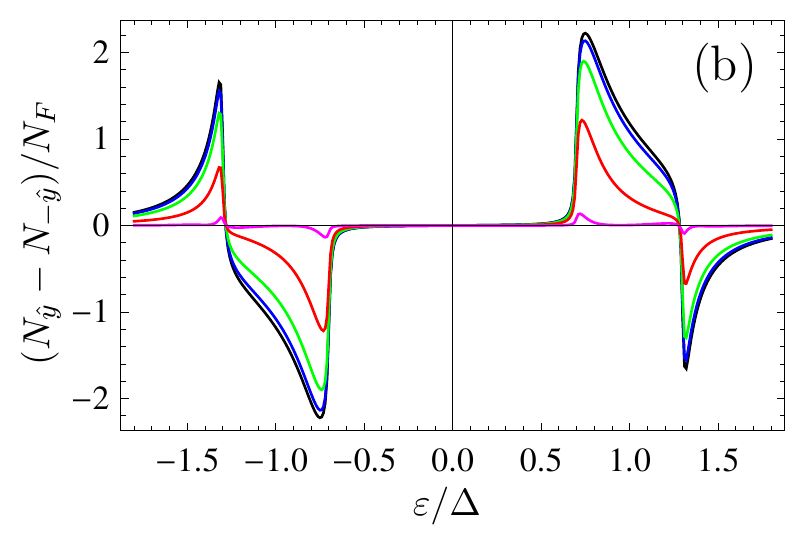}}
   \end{minipage}\hfill
   \begin{minipage}[b]{0.33\linewidth}
   \centerline{\includegraphics[clip=true,width=2.35in]{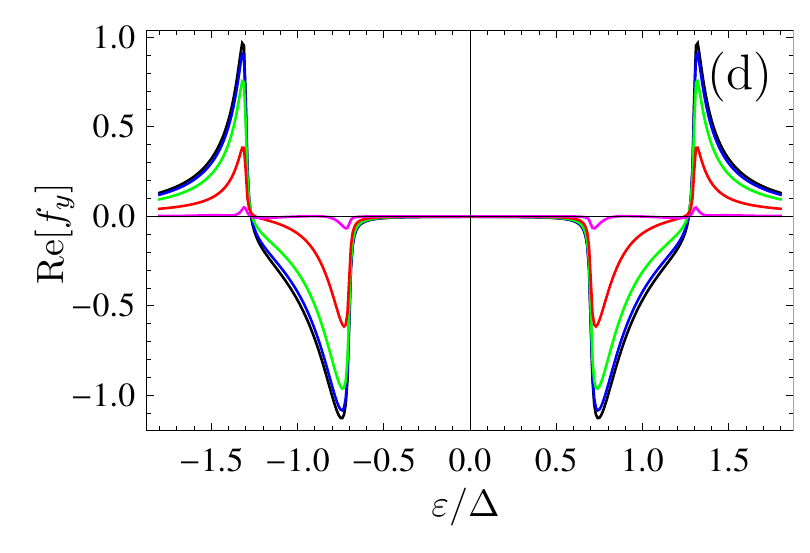}}
   \end{minipage}\hfill
   \begin{minipage}[b]{0.33\linewidth}
   \centerline{\includegraphics[clip=true,width=2.4in]{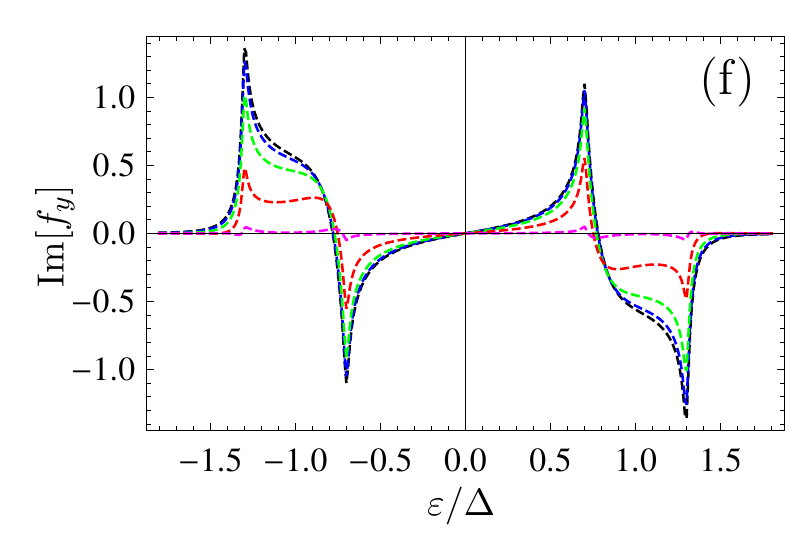}}
   \end{minipage}
   \caption{(a) $N_{\hat x} - N_{-\hat x}$; (b) $N_{\hat y} - N_{-\hat y}$; (c)-(d) ${\rm Re}[\langle f_x\rangle]$ and ${\rm Re}[\langle f_y\rangle]$, respectively; (e)-(f) ${\rm Im}[\langle f_x\rangle]$ and ${\rm Im}[\langle f_y\rangle]$, respectively.  Different curves correspond to different distances $x$ from the wall centre: $x=0$ (black), $x=0.5 \xi_S$ (blue), $x=\xi_S$ (green), $x=2 \xi_S$ (red), $x=5 \xi_S$ (pink). $d_w = \xi_S$. $h_\textrm{eff}=0.3 \Delta$ for all the panels.}
 \label{DOS_f_ballistic}
 \end{figure}

\end{widetext}

Fig.~\ref{DOS_f_ballistic} demonstrates $f_x$ and $f_y$ pairing amplitudes in comparison to the spin-resolved DOS. For this case, we present the results in the fixed spin basis. We see that there is a clear correlation between the real part of $f_{x(y)}$ averaged over trajectories and the corresponding spin-resolved DOS $N_{\hat x(\hat y)} - N_{-\hat x(\hat y)}$. The imaginary part of $f_{x(y)}$ is also strongly correlated with the spin-resolved DOS, but is nonzero at subgap energies, where the corresponding DOS is zero. This fact again illustrates our statement that a nonzero spin-resolved DOS definitely indicates the presence of triplet correlation, while a vanishing spin-resolved DOS does not guarantee their absence.

Far from the DW centre only $f_x$ correlations and $N_{\hat x} - N_{-\hat x}$ are nonzero. This is because far from the wall the magnetization is homogeneous and is directed along the $x$-axis. Therefore only mixed-spin triplet correlations represented by $f_x$ are present in this case. Upon moving to the wall centre the mixed-spin correlations are partially converted into equal-spin ones and therefore the magnitude of $f_y$ grows. For the case of a noncoplanar wall the behavior of $f_z$ is essentially the same as the behavior of $f_y$, while for the case of the coplanar wall $f_z$ is absent.

\section{Conclusions}

We have investigated the influence of inhomogeneous ferromagnetic textures on the structure of the Zeeman-split DOS and the spin-polarized DOS in thin film S/F bilayers. The modifications of the typical Zeeman-split DOS by the spin texture of the ferromagnet show that the spin-polarized DOS clearly indicates the presence and the spatial structure of spin-triplet correlations in the superconducting layer in proximity to a spin-textured ferromagnet. Our results demonstrate the huge potential of tunneling spectroscopic methods in characterizing  superconductivity  in hybrid structures containing magnetic elements. Even more so, by using spin-polarized tunneling information about different spin-component of the local pairing amplitude can be obtained.

\section{Acknowledgement}
This work was financially supported by the DFG through SP 1538 \textit{Spincaloric Transport} and SFB 767 \textit{Controlled Nanosystems}. \textit{Note added:} after finalizing this work a related manuscript was made public \cite{Aikebaier2018}.

 \end{document}